\documentclass[3p,float]{elsarticle}
\usepackage{epsfig}

\begin{document} 


\title{Aspects of a dynamical gluon mass approach to elastic hadron scattering at LHC}

\author[ifgw]{D. A. Fagundes\corref{cor1}} 
\ead{fagundes@ifi.unicamp.br}

\author[ufrgs]{E. G. S. Luna} 
\ead{luna@if.ufrgs.br}

\author[ifgw]{M. J. Menon} 
\ead{menon@ifi.unicamp.br}

\author[ift]{A. A. Natale} 
\ead{natale@ift.unesp.br}

\cortext[cor1]{Corresponding author}

\address[ifgw]{Instituto de F\'{\i}sica Gleb Wataghin, Universidade Estadual de Campinas, UNICAMP, CEP 13083-859, Campinas, SP, Brazil}
\address[ufrgs]{Instituto de F\'{\i}sica, Universidade Federal do Rio Grande do Sul, Caixa Postal 15051, CEP 91501-970, Porto Alegre, RS, Brazil}
\address[ift]{Instituto de F\'{\i}sica Te\'orica, UNESP - Universidade Estadual Paulista, Rua Dr. Bento T. Ferraz, 271, Bloco II
01140-070, S\~ao Paulo - SP, Brazil}




\begin{abstract}
We discuss how the main features of the recent LHC data on elastic scattering can be described by
a QCD-inspired formalism with a dynamical infrared mass scale. For this purpose new developments
on a dynamical gluon mass approach are reported, with emphasis on 
a method to estimate uncertainty bounds in the predictions for the high-energy
scattering observables.
We investigate the effects due to the correlations among the fixed
and free parameters involved and show that the band of predictions are consistent with the recent data
from the TOTEM experiment, including the forward quantities and the differential cross section 
up to the dip position. 
\end{abstract}

\maketitle

\tableofcontents


\section{Introduction}

The TOTEM experiment has been designed to study elastic and diffractive scattering at the LHC, 
providing information on the $pp$ total cross sections and the elastic differential 
cross sections at the energy region 7 - 14 TeV
and therefore expected to represent a crucial source of information for selecting models and pictures. 
First results at 7 TeV have already been published showing that no model predictions
present full agreement with both total cross section and the differential cross section
up to 2.5 GeV$^2$ \cite{totem1,totem2}.

As a consequence reviewing processes are in order in the phenomenological context, and among the variety of
approaches \cite{minijet,others}, the QCD-inspired models certainly play a fundamental role. However, although this class
of models presents explicit connections with minijets and/or semihard concepts, where some kind of perturbative
technique may be applied, the intrinsic soft character involved in the elastic scattering demands a quantitative
connection with the non-perturbative QCD. 

With this aim (see also \cite{giulia01,giulia02,giulia03}),
a novel interpretation for some fundamental parameters in eikonalized QCD-inspired models
has been previously proposed, allowing a quantitative connection between
elastic hadron scattering data and the nonperturbative regime \cite{luna01,luna02,luna03}. This
\textit{Dynamical Gluon Mass} (DGM) approach is based on the possibility that the
nonperturbative dynamics of QCD generates an effective gluon mass at the small momentum transfer
region, which is strongly supported by recent QCD lattice simulations \cite{lttc01}, as well as by 
other phenomenological results
\cite{pheno01,lnz,lns}.
In this context, the increase of the hadronic total cross sections is managed by 
the interplay between the gluon distribution function and a gluon mass scale and
previous analysis on $pp$ and $\bar{p}p$ elastic scattering has indicated good descriptions of the forward data
up to $\sqrt{s}=1.8$ TeV, including the differential cross section at small momentum transfer 
\cite{luna02}.

However,
as commented above, the new results by the TOTEM collaboration demands further tests 
and developments in the phenomenological context and that is the point
we are interested in here. As we shall show and discuss, even not using these new
results as input information for quantitative fit procedures, general aspects of our formalism can be
revisited and improved in the same energy region investigated before, namely
between 10 GeV and 1.8 TeV (some preliminary ideas on these respect
have already been discussed in \cite{lishep}).
Specifically, in this work we present new developments related to some formal and practical
aspects of the approach with focus on: (1) quantitative
account of the energy dependencies in the coupling constant and
dynamical gluon mass; (2) fits based on improved data ensemble;
(3) detailed analysis on the influence in the evaluated quantities associated with
two fundamental parameters, the dynamical gluon mass scale and the soft Pomeron intercept;
(4) estimation of bounds in all evaluated quantities associated with relevant
physical intervals for the above parameters;
(5) explicit reference to some calculational details.
We shall show that these parameters play a crucial role in the energy region
above that used in the fit procedures, which means they can not be fixed at \textit{ad hoc} values
without a fundamental physical justification. 
By considering relevant numerical intervals for the dynamical gluon mass scale and the soft Pomeron intercept,
the question associated with the evaluation
of the corresponding (statistical) uncertainty regions is addressed and a solution based
on bounds for the high-energy predictions is proposed.
Within these uncertainty bands we show that the recent data obtained by the TOTEM
Collaboration at $\sqrt{s}=7$ TeV are reasonably well described, except the differential cross
section beyond the dip position.
Predictions at $\sqrt{s}=14$ TeV and cosmic-ray energy region (Auger) are also presented. The advantages and some
drawbacks associated with this formalism are discussed and further necessary developments outlined
as well.

The text is organized as follows. In Sect. 2 we treat the theoretical context, with explicit
reference to novel formal account of the QCD scale and some calculational aspects. In Sect.
3 the analysis on the influence of the dynamical gluon mass and the soft Pomeron intercept
are presented in detail, as well as the fit procedures, the proposed method
for uncertainty inferences, the predictions at 7 TeV, 14 TeV (LHC), 57 TeV (Auger)
and a discussion on all the obtained results. The conclusions and some
critical remarks are the contents of Sect. 4.

\section{Formalism}

\subsection{Physical Quantities and Eikonal Representation}

Our analysis is based on three physical quantities expressed in terms of the
elastic scattering amplitude $A(s, q)$, where $q = \sqrt{-t}$, with $s$ and $t$ the Mandesltam variables:
the differential cross section
\begin{eqnarray}
\frac{d\sigma}{dq^2}(s,q) = \pi |A(s,q)|^{2},
\end{eqnarray} 
the total cross section
\begin{eqnarray}
\sigma_{\mathrm{tot}}(s) = 4\pi \mathrm{Im} A(s,q=0),
\end{eqnarray} 
and the $\rho$ parameter,
\begin{eqnarray}
\rho(s)=\frac{\mathrm{Re}A(s,q=0)}{\mathrm{Im} A(s,q=0)}.
\end{eqnarray} 
From (1) and (2) the integrated elastic and inelastic cross sections can be obtained,
\begin{eqnarray}
\sigma_{\mathrm{e}l}(s) = \int_{0}^{\infty} \frac{d\sigma}{dq^2}(s,q) dq^2,
\qquad
\sigma_{\mathrm{in}}(s) = \sigma_{\mathrm{tot}}(s) - \sigma_{\mathrm{el}}(s).
\nonumber
\end{eqnarray} 

In the eikonal representation (azimuthal symmetry assumed) the scattering amplitude is
expressed by \cite{pred}
\begin{eqnarray}
A(s,q) = \mathrm{i} \int_{0}^{\infty} b\,db\, J_{0}(qb)\, \{ 1 - e^{\mathrm{i}\,\chi (s,b)} \},
\end{eqnarray}
where $b$ is the impact parameter, 
$J_0$ the zero-order Bessel function and $\chi (s,b)$ the complex eikonal
function in the impact parameter space. Eikonal models can be classified or distinguished according to different
choices for the eikonal either in the impact parameter space or in the momentum transfer space, 
\begin{eqnarray}
\tilde{\chi}(s,q) = \int_{0}^{\infty} b\,db\, J_{0}(qb)\, \chi(s,b),
\end{eqnarray}
where it is connected with the specific dynamical assumptions that characterize each model.
Under the physical condition Im$\,\chi(s,b) \geq 0$ this representation constitutes an 
automatic unitarized framework suitable
for theoretical developments. In special, in the case of QCD based or inspired models the eikonal
in the momentum transfer space is  expressed in terms of hadronic form factors and elementary cross
sections and the main point is how to connect these quantities with
parton-parton scattering processes. Since in these models the rise of the total cross section at high energies is 
driven by gluon-gluon semihard scattering, this is the main process to be investigated.

\subsection{Dynamical Gluon Mass and Gluon-Gluon Cross Section}

In addition to lattice QCD \cite{lttc01}, the Schwinger-Dyson equations constitutes a fundamental nonperturbative
formalism in the investigation of hadron physics, specially in what concerns confinement
and dynamical chiral symmetry breaking. Infrared finite solutions for these equations
have been obtained by Cornwall by means of the pinch technique
leading to a gluon propagator which carries a dynamical mass \cite{cornwall1,cornwall2,cornwall3}.
A review on this dynamical gluon mass generation is presented in \cite{alkofer},
discussions and more details can be found in \cite{luna02} and references therein.
Here we display the main formulas connected with the nonperturbative
gluon-gluon cross section, stressing a formal improvement regarding the 
previous work \cite{luna02}. 

In the DGM approach the nonperturbative integrated elastic cross section
for the scattering $g g \rightarrow g g$, expressed in terms of
the subprocess energy $\hat{s}$, is given by \cite{luna02}
\begin{eqnarray}\label{eq:crosssec_npt_gg}
\hat{\sigma}_{gg}(\hat{s}) = \left(\frac{3\pi \bar{\alpha}_{s}^{2}}{\hat{s}}\right) \left[\frac{12\hat{s}^{4}
- 55 M_{g}^{2} \hat{s}^{3} + 12 M_{g}^{4} \hat{s}^{2} + 66 M_{g}^{6} \hat{s} -
8 M_{g}^{8}}{4 M_{g}^{2} \hat{s} [\hat{s} - M_{g}^{2}]^{2}} 
- 3 \ln \left( \frac{\hat{s} - 3M_{g}^{2}}{M_{g}^{2}}\right)\right],
\label{h16}
\end{eqnarray}
where $\bar{\alpha}_{s} = \bar{\alpha}_{s}(\hat{s})$ and $M_{g} = M_{g}(\hat{s})$ are the
running coupling constant and the dynamical gluon mass, respectively. From the Cornwall solution
for the gluon propagator in the case of pure gauge QCD, these quantities
are expressed by \cite{cornwall1,cornwall2,cornwall3}
\begin{eqnarray}
\bar{\alpha}_{s} (\hat{s})= \frac{4\pi}{\beta_0 \ln\left[
(\hat{s}+4M_g^2(\hat{s}))/\Lambda^2 \right]},
\label{acor}
\end{eqnarray}

\begin{eqnarray}
M_g^2(\hat{s}) = m_g^2 \left[ \frac{\ln \left( \frac{\hat{s}+4{m_g}^2}{\Lambda^2} \right) }{\ln
\left( \frac{4m_g^2}{\Lambda^2} \right) } \right]^{- 12/11} ,
\label{mdyna}
\end{eqnarray}
where $\beta_{0} = 11 - \frac{2}{3}n_{f}$ ($n_{f}$ is the number of flavors), $\Lambda = \Lambda_{QCD}$ and 
$m_{g} \equiv M_g(0)$ is the gluon mass scale. 
A fundamental consequence of this result is its extension to perturbative QCD. In fact, in the
limit $\hat{s} \gg \Lambda^2$ we have $M_g^2(\hat{s}) \rightarrow$ 0 and  $\bar{\alpha}_{s}$
matches the one-loop perturbative QCD one \cite{luna02}.

In this context, by considering $n_{f}$ = 4 and $\Lambda$ = 284 MeV, the only unknown quantity in the
evaluation of  $\hat{\sigma}_{gg}(\hat{s})$ is the gluon mass scale $m_g$,
the IR value for $M^2_g(\hat{s})$.
Phenomenological analyses suggest typical values of $m_g$ in the range from 300 MeV
up to 700 MeV \cite{cornwall1,cornwall2,cornwall3,amn2,an,pheno01,lnz,lns}
and in \cite{luna02} it has been inferred $m_g =$ 400 MeV. To investigate the relevant numerical
interval for $m_g$ and the effects in the analysis of elastic hadron scattering is one
of the main points we are interested in here.

It should be noted that in previous works \cite{luna01,luna02,luna03,lnz} the dependence of $M^2_g$ on the energy
of the subprocess $g g \rightarrow g g$ has been neglected as a first approximation, that is, in Eq. (\ref{h16}) the IR value
$m^2_g$ has been used in place of $M^2_g(\hat{s})$. Therefore, the integrated cross section as given by Eq. (6)
generalizes the one considered in \cite{luna01,luna02,lnz}, enlarging the physical meaning of the dynamical gluon mass
and the efficiency of the formalism in the data analyses, as shown in Sect. 3.
Another parametrization for $M^2_g(\hat{s})$ is discussed in Appendix A.

\subsection{Eikonalized Approach}

At high energies the soft and the semihard components of the scattering amplitude are closely related
\cite{gribov,ryskin01}, and it becomes important to distinguish between semihard gluons, which participate in hard
parton-parton scattering, and soft gluons, emitted in any given parton-parton QCD radiation process. Hence a formalism
based on QCD has to incorporate soft and semihard processes in the treatment of high-energy hadronic interactions 
and, more importantly, has to bring up information about the infrared properties of QCD. 

In the phenomenological context, the developments of the mini-jet or semi-hard QCD picture \cite{minijet}
constitute an interesting step in the
search for connections between
hadron scattering and QCD subprocesses, associated with 
gluon-gluon ($gg$), quark-gluon ($qg$) and quark-quark ($qq$) interactions. That is the case in 
the formalisms
by Margolis \textit{et al}. \cite{qcdi1,qcdi2,qcdi3} and the more recent
version sometimes referred to as Aspen model, by Block \textit{et al}. \cite{block1,block2}.
However, despite its efficiency in the description 
of the forward quantities in the elastic channel,
as well as differential cross sections at small values of the momentum transfer
a serious drawback with the Aspen model concerns the absence of the expected connections with
nonperturbative QCD. 
Moreover, and as a consequence, two fundamental quantities, the infrared mass scale,
denoted  $m_0$, and the coupling constant $\alpha_s$, are unknown parameters fixed
to \textit{ad hoc} values 600 MeV and 0.5 respectively, in order
to obtain best fits in data analyses \cite{block1,block2}.

In this respect, the main ingredient in the DGM approach \cite{luna01,luna02} has been
the novel physically motivated interpretation for both quantities
through the gluon mass scale as a natural
regulator of the infrared divergences associated with the semihard gluon-gluon subprocess cross sections,
leading therefore to direct connections with nonperturbative QCD.
The general formulation is largely based on the eikonal structure of the Aspen model \cite{block1,block2},
but with the above well founded connection, as well as differences regarding free/fixed parameters,
constraints and specific parametrization.
The point is to consider Eqs. (6 - 8) for the $gg$ subprocess and the
IR values $\bar{\alpha}_{s} (\hat{s}=0)$ and
$M^2_g(\hat{s}=0) = m_g^2$ in substitution to $\alpha_s$ and $m_0$ in all elementary subprocesses.

Specifically, for $pp$ and $\bar{p}p$ elastic scattering, the complex eikonal in the
impact parameter space is expressed in terms of even ($e$) and odd ($o$) functions,
\begin{equation}
\chi_{\bar{p}p}(s,b) = \chi_{e} (s,b) + \chi_{o} (s,b),
\qquad
\chi_{pp}(s,b) = \chi_{e} (s,b) - \chi_{o} (s,b).
\end{equation}

The odd part is expected to account for the difference between both reactions
at low energies and to vanish in the high-energy limit. It is therefore parametrized
as a Reggeon contribution and in a factorized form,
with the inclusion of the dynamical gluon mass ingredients and the odd prescription,
\begin{eqnarray}
\chi_{o} (s,b) = k\, C_{o}\,
\frac{m_{g}}{\sqrt{s}} \, \mathrm{e}^{i\pi /4}\, 
W(b;\mu_{o}),
\label{bl2}
\end{eqnarray}
where 
\begin{eqnarray}
k \equiv \frac{9\pi \bar{\alpha}^2_{s}(0)}{m_{g}^2},
\end{eqnarray}
$W(b;\mu_{o})$ is the Fourier transform of the dipole form factor
\begin{eqnarray}
W(b;\mu_{o}) = \frac{\mu_{o}^2}{96\pi}[\mu_{o}b]^3 K_3(\mu_{o}b)
\end{eqnarray}
($K_3$ is the modified Bessel function of second kind), and $\mu_{o}$ and $C_o$ are free fit parameters.

The even part is supposed to be expressed in terms of contributions from elementary interactions, 
namely gluon-gluon, quark-gluon and quark-quark, in a factorized form at the constituent level,
\begin{eqnarray}
\chi_{e}(b,s) &=& \chi_{qq} (b,s) +\chi_{qg} (b,s) + \chi_{gg} (b,s) \nonumber \\
&=& \mathrm{i} [\sigma_{qq}(s) W(b;\mu_{qq}) + \sigma_{qg}(s) W(b;\mu_{qg})
+ \sigma_{gg}(s) W(b;\mu_{gg})] ,
\label{fatoriz1}
\end{eqnarray}
where, for $ij = qq, qg, gg$, $W(b,\mu_{ij})$ are overlap functions with
the same structure of Eq. (12), $\mu_{ij}$  are free fit parameters and $\sigma_{ij}$ are the corresponding elementary
cross sections.

For the sub-processes involving quarks, the parametrizations for the cross sections 
follows from approximate forms of distribution functions
involving quarks and gluons at small $x$ region, with the even prescription 
in the corresponding eikonal:
\begin{eqnarray}
\sigma_{qq}(s) = k \, C_{qq} \, \frac{m_{g}}{\sqrt{s}}\, \textrm{e}^{i\pi/4}, 
\label{mdg1}
\end{eqnarray}
\begin{eqnarray}
\sigma_{qg}(s) = k \left\lbrace C_{qg} + C'_{qg} \left[ \ln{\left(\frac{s}{m_{g}^{2}}\right)}-i\frac{\pi}{2} \right] \right\rbrace,
 \label{mdg11}
\end{eqnarray}
where $C_{qq}$, $C_{qg}$,  $C_{qg}^{\prime}$, are free fit parameters. The convenience of the general structure adopted
in the parametrizations
(\ref{bl2}), (\ref{fatoriz1}), (\ref{mdg1}) and (\ref{mdg11}) can be traced back to the early work of Margolis \textit{et al.}  \cite{qcdi1,qcdi2,qcdi3} and references therein.

The full $gg$ cross section  is expressed by
\begin{eqnarray}
\sigma_{gg}(s) = C_{gg} \int_{4m_{g}^{2}/s}^{1} d\tau \,F_{gg}(\tau)\,
\hat{\sigma}_{gg} (\hat{s}) ,
\label{sloh1}
\end{eqnarray}
where $\tau = \hat{s}/s$, $F_{gg}(\tau)$ is the convoluted gluon distribution function,
\begin{eqnarray}
F_{gg}(\tau)=[g\otimes g](\tau)=\int_{\tau}^{1} \frac{dx}{x}\, g(x)\,
g\left( \frac{\tau}{x}\right),\label{fgg}
\end{eqnarray}
and $\hat{\sigma}_{gg}(\hat{s})$ represents the subprocess $gg\rightarrow gg$ cross section,
given here by the nonperturbative result (6). For the gluon distribution function,
since in the DGM approach the small-$x$ semihard gluons play a central role, the phenomenological 
parametrization used in \cite{block1} is also considered here, 
\begin{eqnarray}
g(x) = N_{g} \, \frac{(1-x)^5}{x^{J}},
\label{distgf}
\end{eqnarray}
which simulates the effect of scaling violations in the small $x$. Here $J$ can be associated with the soft
Pomeron intercept ($J=1+\epsilon$). Notice that the ansatz (\ref{distgf}) satisfies the normalization
condition $\int_{0}^{1} dx\, x g(x)=A $, where $A$ is a constant \cite{georgi01}.
If $A=1/2$, $N_{g} = (6-\epsilon)(5-\epsilon)...(1-\epsilon)/240 $. The power $5$ appearing in the term $(1-x)$ is
suggested by dimensional counting rules \cite{georgi01,owens01,cutler01}.

The integration of Eq. (16), taking  account of the analyticity properties, demands some comments.
The odd contribution to the eikonal, Eq. (10) and those from $qg$ and $qq$ subprocesses to the even
eikonal, Eqs. (14) and (15), are closed analytical forms, so that the corresponding prescriptions
are straightforward to be applied, leading to the generation of the real and imaginary parts. 
On the other hand, the $gg$ contribution, Eq. (16), cannot be analytically
evaluated and any numerical approach, embodying analyticity, demands some assumptions.
We have used two independent methods. In the first one generated numerical points for the 
real primitive of the integrand in (16) have been parametrized by polynomials in $\ln s$ and then the even prescription
has been applied. In the second one we have used first order derivative dispersion
relation for even functions \cite{am} applied as
\begin{eqnarray}
\mathrm{Re}\, \sigma_{gg}(s) \approx \frac{\pi}{2} \frac{d}{d \ln{s}}\, \mathrm{Im}\, \sigma_{gg}(s),
\end{eqnarray}
and numerical derivation of the generated points. In (19), 
$\mathrm{Im}\, \sigma_{gg}(s)$ corresponds to the generated numerical points for
the real primitive of (16).
Both methods lead to the same results, except for negligible differences at low energies.

With all the above ingredients the formalism is complete. Specifically, it has two fundamental
parameters, the \textit{dynamical gluon mass scale} $m_g$ and the 
\textit{soft Pomeron intercept} $\epsilon$. Once $m_g$ and $\epsilon$ are chosen, the model has
9 free parameters to be determined by fits to the experimental data: 5 normalization constants associated with
the odd contribution and elementary cross section ($C_o$, $C_{qq}$, $C_{qg}$, $C'_{qg}$, $C_{gg}$) 
and 4 coming from the corresponding form factors ($\mu_{o}$, $\mu_{qq}$, $\mu_{qg}$, $\mu_{gg}$).
That should be contrasted with
\textit{6 fixed} and 6 fit parameters in the Aspen model \cite{block1,block2}.

In what concerns the momentum transfer space, the essential  ingredient in the structures of the eikonals (9), (10) and (13) is the 
combination of four \textit{dipole} form factors, of type (12) in the impact parameter space (in fact, a linear combination at each 
fixed energy).  Therefore, the dependence of the physical quantities on the momentum transfer is a direct consequence of these choices 
and in this respect the following comments are in order. Historically, the \textit{dipole ansatz} may be traced back to the conjecture
by Wu and Yang that the distribution of hadronic matter inside a hadron is similar in electromagnetic and strong interactions 
\cite{wuyang}. Specifically they conjectured that the elastic $pp$ differential cross section might be proportional to the fourth 
power of the \textit{proton charge form factor}, that is the form factor measured in electron-proton scattering. In the context of the 
Chou-Yang model \cite{chouyang1}, with an input for the form factor, the differential cross section can be evaluated. The use of a 
parametrization as a sum of exponential in $q^2$ by Chou and Yang \cite{chouyang2} and the dipole electric form factor by Durand and 
Lipes \cite{durandlipes},
\begin{eqnarray}
G_D(q) = \frac{1}{[1+q^2/\mu^2]^2},
\ \ \ 
\mu^2 = 0.71\ \textnormal{GeV}^2,
\nonumber
\end{eqnarray}
led to the prediction of the diffractive pattern in $pp$ differential cross section and, more importantly, the correct position of the dip,
as experimentally observed latter. However, the dipole parametrization predicts multiple dips and peaks that have never been observed in
the experimental data. In our case (as well as in the Aspen model), the oscillatory behaviour, characteristic of the dipole structure can 
be attenuated by the combination of four different dipoles, at least up to the intermediate region of the momentum transfer 
(dip-bump structure). However, beyond this region, changes of slopes are expected, as indicated by the results presented in the following 
section. This effect seems not to be present in the published plots by the TOTEM Collaboration \cite{totem1}, suggesting that other form factors should be
regarded in order to describe the high-$t$ data. We shall return to this point in the conclusions and final remarks.

\section{Testing parameters: Fits and Results}

 Despite the suitable formal connection with nonperturbative QCD and the efficiency in the description of the experimental
data, one important aspect to be investigated in the DGM approach concerns the influence of the fundamental parameters in the evaluated quantities (fits and predictions): $m_g$ and $\epsilon$.
They could certainly be fixed, as done for example with the corresponding parameters 
(and others) in the Aspen model \cite{block1}. However, as we shall show, relevant numerical intervals for $m_g$
and $\epsilon$ affect the model predictions at the highest energies and that puts serious limits on the reliability of models with \textit{ad hoc} values for the fixed parameters involved. Our goal here is to point out the relevance of this aspect and the fact that it demands some solution or a clear explanation, prior to consider these models as physically well founded.
To this end we first discuss some aspects of the data ensemble used here (which differs from the one considered in \cite{luna01,luna02}) and outline the fit procedure (Subsection 3.1). Then we treat in some detail the selection of the relevant intervals for the parameters and the proposed methodology for uncertainty inferences (Subsection 3.2), followed by a critical discussion on the obtained results (Subsection 3.3).

\subsection{Data Ensemble and Fit Procedure}

In the previous work \cite{luna02} the data ensemble used in the fits includes only forward
quantities, $\sigma_{tot}$, $\rho$ parameter and the slope parameter $B$ up to
1.8 TeV. Here we shall consider a different ensemble for the reasons that follows.

Even in the very forward direction the evaluation of the slope parameter depends somehow
on the momentum transfer interval considered (at least two points), which renders difficult comparison among
results from different experiments with different intervals. That is clearly illustrated in the comparison of the
recent result by the TOTEM Collaboration with the existing data  \cite{totemslope} that includes
the pp2pp result at 200 GeV \cite{pp2pp}. Moreover, in a model context the slope is evaluated
at one point, $q^2$ = 0, and therefore some bias may always be present in the calculation. For that reason, we
shall not consider here the forward slope as input quantity in the data reductions.

On the other hand, as constructed, the formalism is expected to be applicable in the soft region, which in
terms of the differential cross section means intervals in momentum transfer roughly up to 1 GeV$^2$,
or even beyond. The presence of four dipole form factors in the eikonal
(Subsection 2.3) corroborates this expectation. Hence we include in the ensemble to be fitted
the differential cross section data available at the highest energies,
namely $\bar{p}p$ scattering at 546 GeV and 1.8 TeV,
covering the region up to $q^2 \sim$ 1.5 GeV$^2$ and $\sim$ 0.6 GeV$^2$, respectively.

Since the numerical tables on the differential cross section at 7 TeV are not yet available
and a second measurement of the total cross section (through a luminosity independent
method) is expected, we shall not consider here the recent TOTEM results as input in the
fits. However, the numerical information available (and published) will be referred to
as a guidance in the discussion of our predictions.

Our ensemble therefore includes the $pp$ and $\bar{p}p$ data on $\sigma_{tot}$ and $\rho$ above 10 GeV
up to 1.8 TeV, together with the $\bar{p}p$ differential cross section data at 546 GeV and 1.8 TeV. The data
have been collected from the PDG \cite{pdg} and Durham \cite{durham} sites and in all cases 
the statistic and systematic errors have been added in quadrature.
The data reductions consist in assuming some input values for m$_{g}$ and $\epsilon$ and through Eqs. (1) to (3)
to fit the above data ensemble.
For the fit program we have used the class TMinuit of the CERN ROOT Framework \cite{root} and the MIGRAD
minimizer, setting the confidence level at 90\%.

\subsection{Testing Parameters}

Preliminary results with the above ensemble and fit procedure have already been presented in our
previous analysis \cite{lishep}. The main conclusion has been that by considering narrow intervals for
$m_g$ and $\epsilon$ all the
fitted data are quite well described in all cases. However, at higher energies the predictions are
sensitive to each numerical interval considered.
Here, we address the question on
how to infer quantitative uncertainties in the model predictions taking into account relevant intervals
for $m_g$ and $\epsilon$. It is important to stress that our point does not concern standard 
propagation from the errors in the fit parameters, as given by the error matrix (variances and covariances).
Our focus is only on the effects of physically motivated intervals for $m_g$ and $\epsilon$.

However that is not an easy task,  due to  the numerical
integrations involved, the absence in this case of a statistical method to extract uncertainties and the strong
correlations among $m_g$, $\epsilon$ and all the free fit parameters, as we shall shown. What we propose here concerns a method to infer \textit{bounds} in the evaluated quantities in accordance
with relevant intervals for $m_g$ and $\epsilon$, in a similar way as has been recently done by Achilli
\textit{et al}. in the context of minijet models \cite{achilli}.

Due to the correlations among $\epsilon$, $m_g$ and the fit parameters, a relevant interval
for one of the fundamental parameters depends on the relevant interval for the other one. For that reason we first
treat $\epsilon$ and then $m_g$.
For the soft Pomeron intercept, as in \cite{lishep}, we consider the results from the detailed analyses on the
extrema and constrained bounds for the intercept, since they are based on both scattering data \cite{lm}
and scattering together with spectroscopy data \cite{lmm}, respectively. With these bounds,
namely $\epsilon_{lower}$ = 0.080 and $\epsilon_{upper}$ = 0.090 \cite{lm,lmm}, we consider as relevant interval
all real values in this range, including the bounds. Thus we use the following notation for intervals with that
meaning (equally likely  values):
\begin{eqnarray}
\epsilon: [0.080,\ 0.090].
\nonumber
\end{eqnarray}
For the dynamical gluon mass scale, as commented in Subsection 2.2,
phenomenological analyses suggest $m_g$ values in the range from 300 MeV up to 700 MeV.
Taking into account the interval for $\epsilon$, as we shall show, the following interval
for $m_g$, larger than that considered in \cite{lishep}, and with the above notation
and meaning, is adequate to our purposes:
\begin{eqnarray}
m_g (\mathrm{MeV}): [300,\ 600].
\nonumber
\end{eqnarray}

Once selected the relevant intervals, the next step is to investigate their effects
in the fit results and evaluated quantities, looking for the corresponding uncertainties
in these quantities.
At this point it is important to stress that, in principle, any
numerical pair ($\epsilon$, $m_g$) in the selected intervals is equally likely and therefore
we can not attribute central values with uncertainty, as for example
0.085 $\pm$ 0.005 for $\epsilon$ and 450 $\pm$ 150 for $m_g$. That is, there is no meaning in assuming that
a most probable value is any closer to the middle
of this range that it is to either endpoint.
Moreover, it is also not possible to test
the effects of all real pairs ($\epsilon$, $m_g$). For those reasons we have considered a method
consisting of the following steps:

\begin{itemize}
\item[1.]
From different tests with real pairs ($\epsilon$, $m_g$) we have selected relevant values for
each parameters in each interval, with the result:
\begin{eqnarray}
\epsilon &:& 0.080,\, 0.085,\, 0.090 \nonumber \\
m_g &:& 300,\, 400,\, 500,\, 600 \, \, \mathrm{MeV} 
\nonumber
\end{eqnarray}
\item[2.]
For each of the above pairs ($\epsilon$, $m_g$), therefore 12 variants, we have performed the
data reductions (Subsection 3.1), determining in each case the $\chi^2$ per degrees of
freedom (DOF) as a test for goodness of fit.
\item[3.]
Based on the above statistical information and in the global description of the
fitted data, we have determined an optimum value for one of the parameters together
with the relevant interval for the other one. 
\item[4.]
By fixing one parameter in its optimum value the extrema values of the interval
corresponding to the other parameter have been used as upper and lower bounds in
the evaluated quantities.
\item[5.]
The band limited by the above upper and lower bounds has been considered as
an estimation of the uncertainty region in the evaluated quantities.
\end{itemize}

With this strategy we do not take account of the simultaneous influences of both intervals,
but only the effect of the interval in one parameter for a fixed value of the other parameter.
We shall return to this point in subsection C.

\begin{table}[ht]
\caption{Values of the $\chi^{2}$/DOF for 318 DOF obtained in all fits.}\label{tab1}
\begin{center}
\begin{tabular}{c|cccc}
\hline
            &        & $m_g$   & MeV    &       \\
 $\epsilon$ & 300    & 400     & 500    & 600   \\
\hline
0.080       & 1.160  & 0.9487  & 0.9690 & 0.9984 \\
0.085       & 0.9899 & 0.9536  & 0.9536 & 1.432  \\
0.090       & 1.001  & 0.9597  & 1.070  & 4.339  \\
\hline
\end{tabular}
\end{center}
\end{table}
\begin{figure}[h]
\begin{center}
\includegraphics[width=15cm,height=10cm]{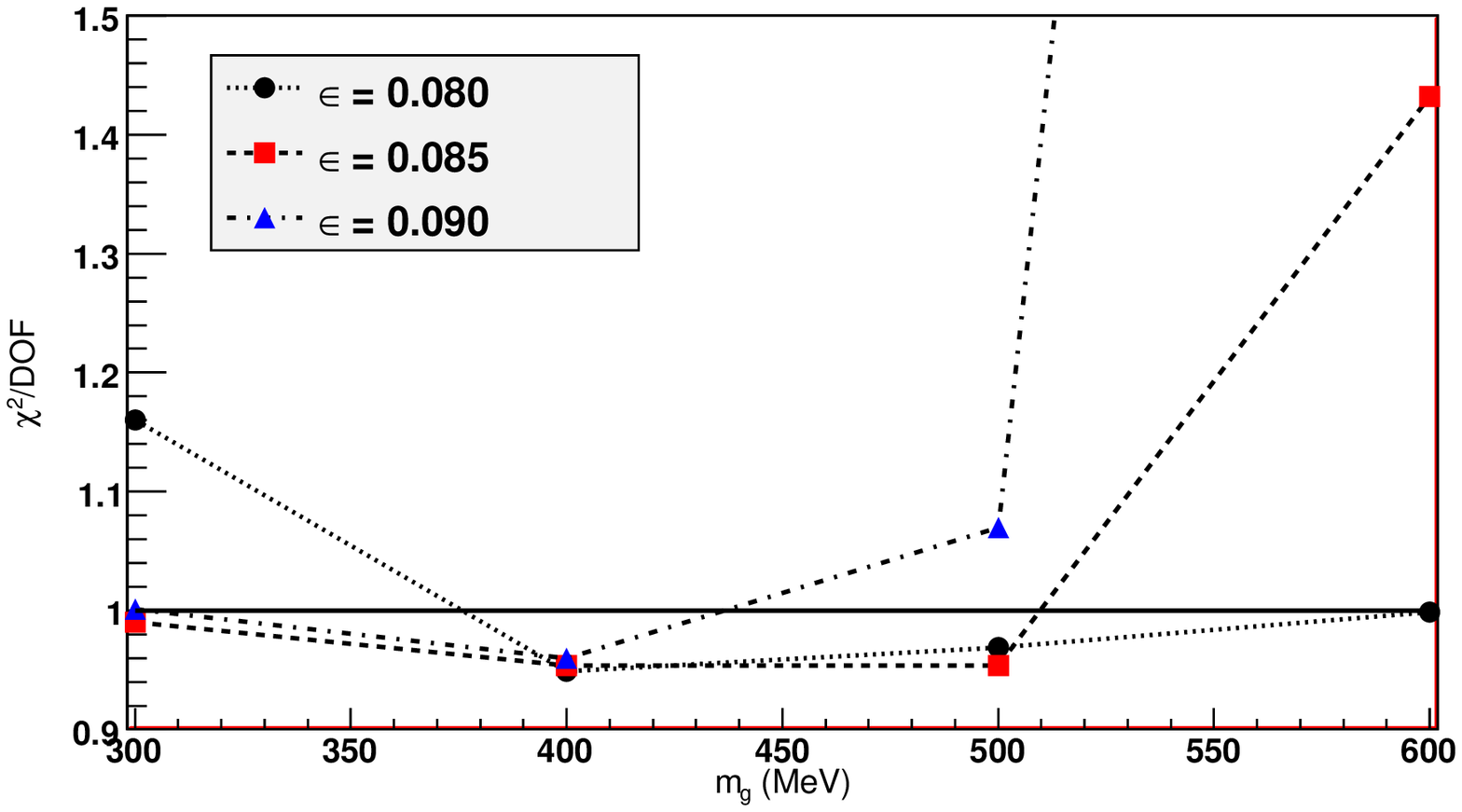}
\end{center}
\caption{Reduced chi-square in terms of the dynamical gluon mass scale for different
values of the soft Pomeron intercept. The straight line segments are drawn only to guide the eyes.}
\label{f1}
\end{figure}

The values of the $\chi^2$/DOF for each one of the 12 variants are displayed in Table
I and plotted in Fig. 1, in terms of the dynamical gluon mass scale and for each value
of the soft Pomeron intercept. Since we have 318 DOF the best results demand 
$\chi^2$/DOF closest and/or below 1. Figure 1 plays a central role in our analysis
since we can extract the following main conclusions:

\begin{itemize}
\item[1.]
For $m_g$ = 600 MeV the cases $\epsilon$ = 0.085 and 0.090 are excluded on statistical grounds.
\item[2.]
For $\epsilon$ = 0.080, despite the reasonable result at $m_g$ = 300 MeV, the whole interval
in $m_g$ corresponds to quite good statistical results.
\item[3.]
For $m_g$ = 400 MeV the whole interval for $\epsilon$ corresponds to quite good statistical results.
\item[4.]
For both parameters considered simultaneously the intervals $\epsilon$: [0.080, 0.090] and
$m_g$ (MeV): [400,$\,$ 500] correspond to acceptable statistical results.
\end{itemize}

Despite the last conclusion, in order to develop here a feasible and economic method for
uncertainty inferences, we do not consider both intervals simultaneously. For individual
intervals, based on the first three statistical conclusions and after checking the descriptions of the
fitted data in all cases analyzed, we propose the following optimum  solutions, with the interval notation
and meaning defined before:

\vspace{0.3cm}

- For fixed $\epsilon$ = 0.080 $\rightarrow$ relevant interval $m_g$ (MeV): [300,$\,$ 600];

\vspace{0.3cm}

- For fixed $m_g$ = 400 MeV $\rightarrow$ relevant interval $\epsilon$: [0.080,$\,$ 0.090].

\vspace{0.3cm}

We note that the asymmetrical character of this solution is a consequence of the statistical analysis 
we have considered.
Moreover, as we show in what follows, this solution allows us to estimate the uncertainties
in the evaluated quantities in a consistent way and that is the point we are interested in here.

\begin{table}[ht]
\caption{Values of the fit parameters for $\epsilon = 0.080$. $C_o$, $C_{qq}$, $C_{qg}$, $C'_{qg}$ and $C_{gg}$ are dimensionless 
and  $\mu_{o}$, $\mu_{qq}$, $\mu_{qg}$, $\mu_{gg}$ are in GeV.}
\begin{center}
\begin{tabular}{ccccc}
\hline
$m_{g}\ (\textrm{MeV}) : $ & 300 & 400 & 500& 600\\
\hline
$C_{o}$ & 1.170$\pm$0.026& 3.03$\pm$0.40 & 5.61$\pm$0.24&10.12$\pm$0.41\\
$C_{qq}$ & 3.8498$\pm$0.0036  & 10.7$\pm$1.4&30.65$\pm$0.72& 21.2$\pm$1.9\\
$C_{qg}$($\times$10$^{-1}$) & 1.8993$\pm$0.0082 & 8.74$\pm$0.59 & 13.414 $\pm$0.034& 42.70$\pm$0.62\\
$C'_{qg}$($\times$10$^{-2}$) & 1.8174$\pm$0.0068 & 4.51$\pm$0.62 & 18.842$\pm$0.060&18.55$\pm$0.74\\
$C_{gg}$($\times$10$^{-3}$) & 0.979$\pm$0.019 & 3.79$\pm$0.17 & 6.832$\pm$0.057&18.20$\pm$0.26\\
$\mu_{o}$ & 0.1862$\pm$0.0041 & 0.41$\pm$0.17 & 0.196$\pm$0.021&0.622$\pm$0.053\\
$\mu_{gg}$ & 0.6463 $\pm$0.0055 & 0.651$\pm$0.066 & 0.6314$\pm$0.0031&0.6427$\pm$0.0020\\
$\mu_{qq}$ & 0.156218$\pm$0.0023& 1.32$\pm$0.16&1.574$\pm$0.047&1.03$\pm$ 0.15\\
$\mu_{qg}$ & 0.8368$\pm$0.0027 & 0.838$\pm$0.044 & 0.8214$\pm$0.0020&0.8377$\pm$0.0018\\
\hline

\end{tabular}
\end{center} 
\end{table}
\begin{figure}[pb]
\centering
\epsfig{file=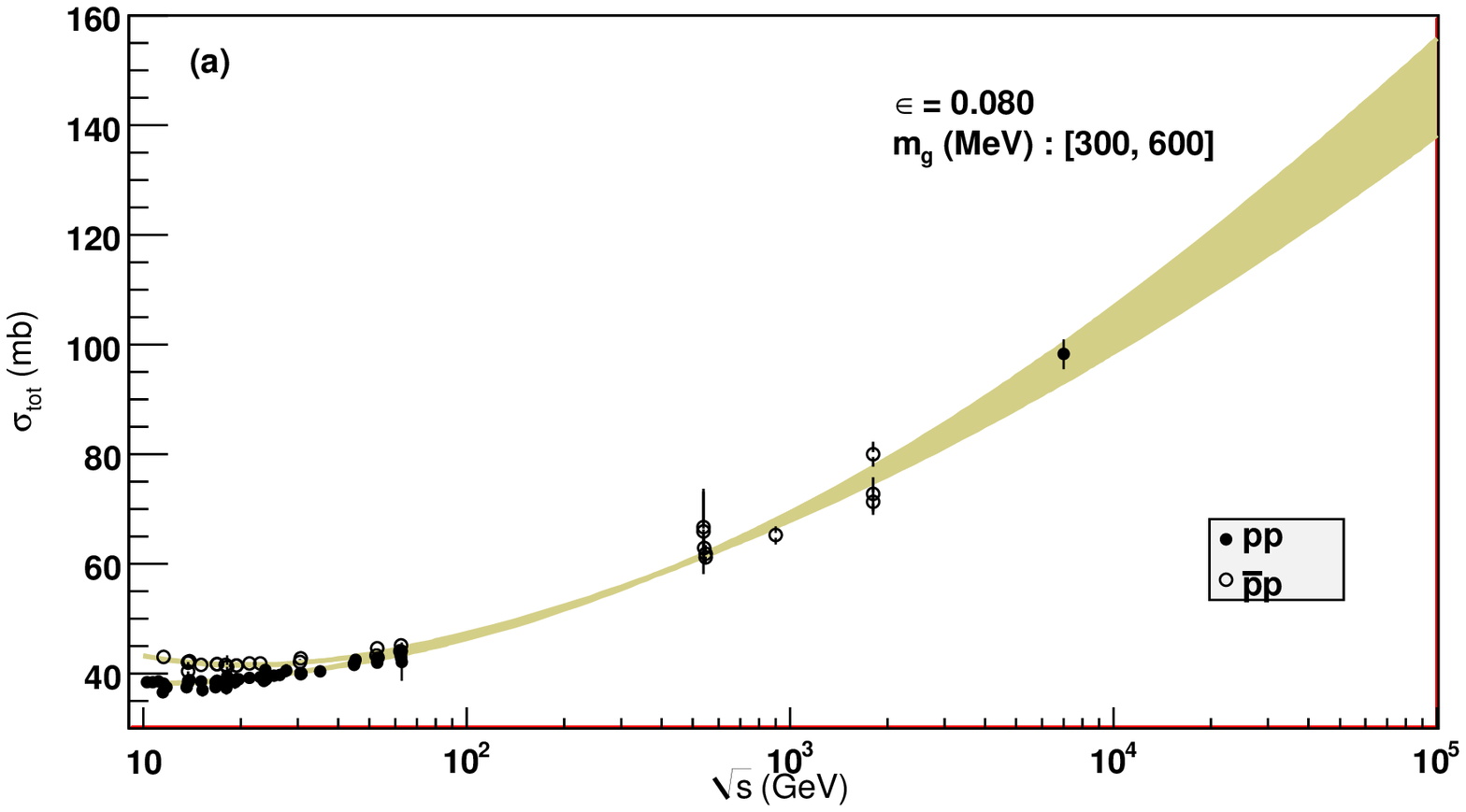,width=7.6cm,height=8cm}
\epsfig{file=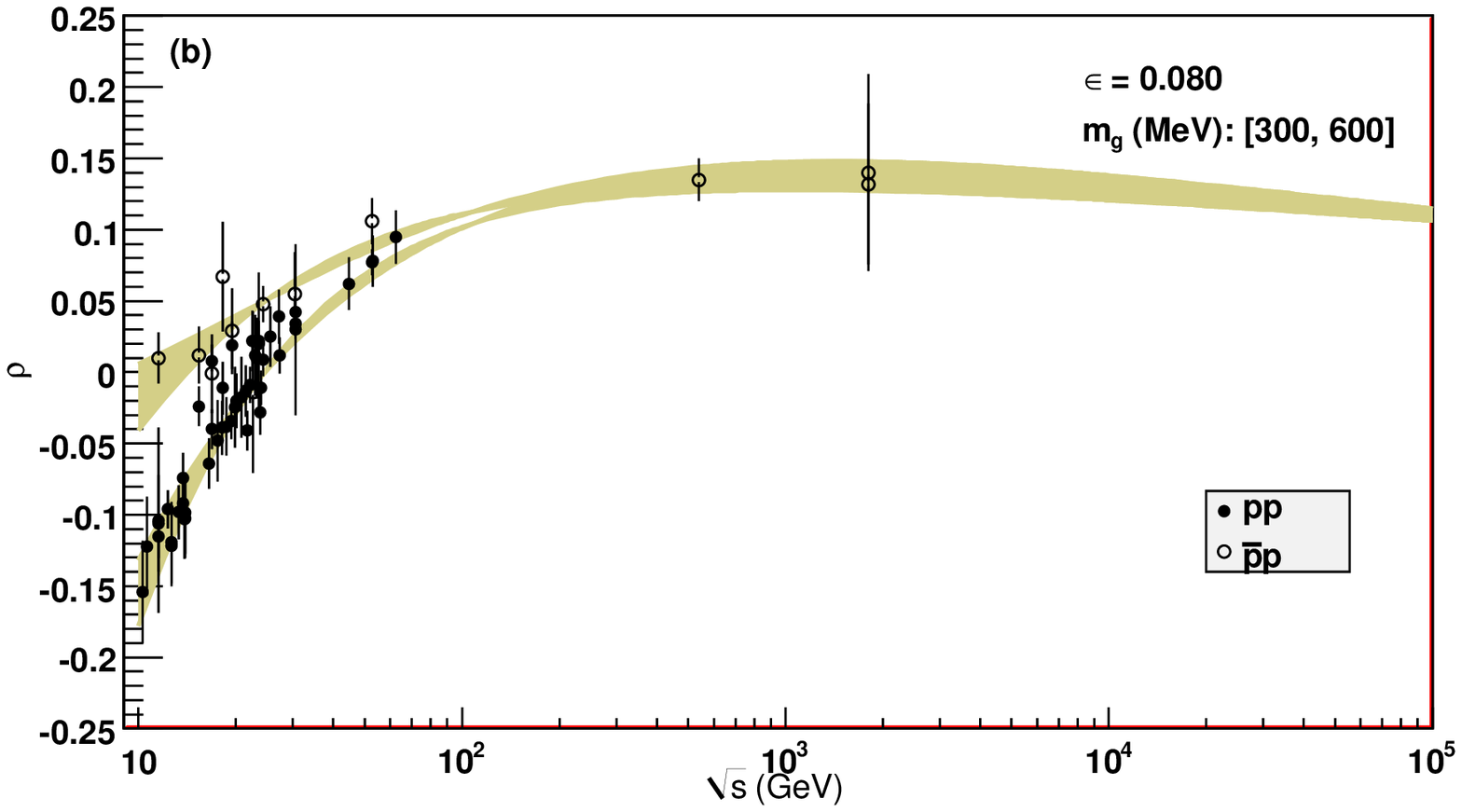,width=7.6cm,height=8cm}
\caption{Total cross sections and $\rho$ parameter for fixed $\epsilon$ = 0.080, upper bound for $m_g$ = 600 MeV
and lower bound for  $m_g$ = 300 MeV.}
\label{f2}
\end{figure}
\begin{figure}[pb]
\centering
\epsfig{file=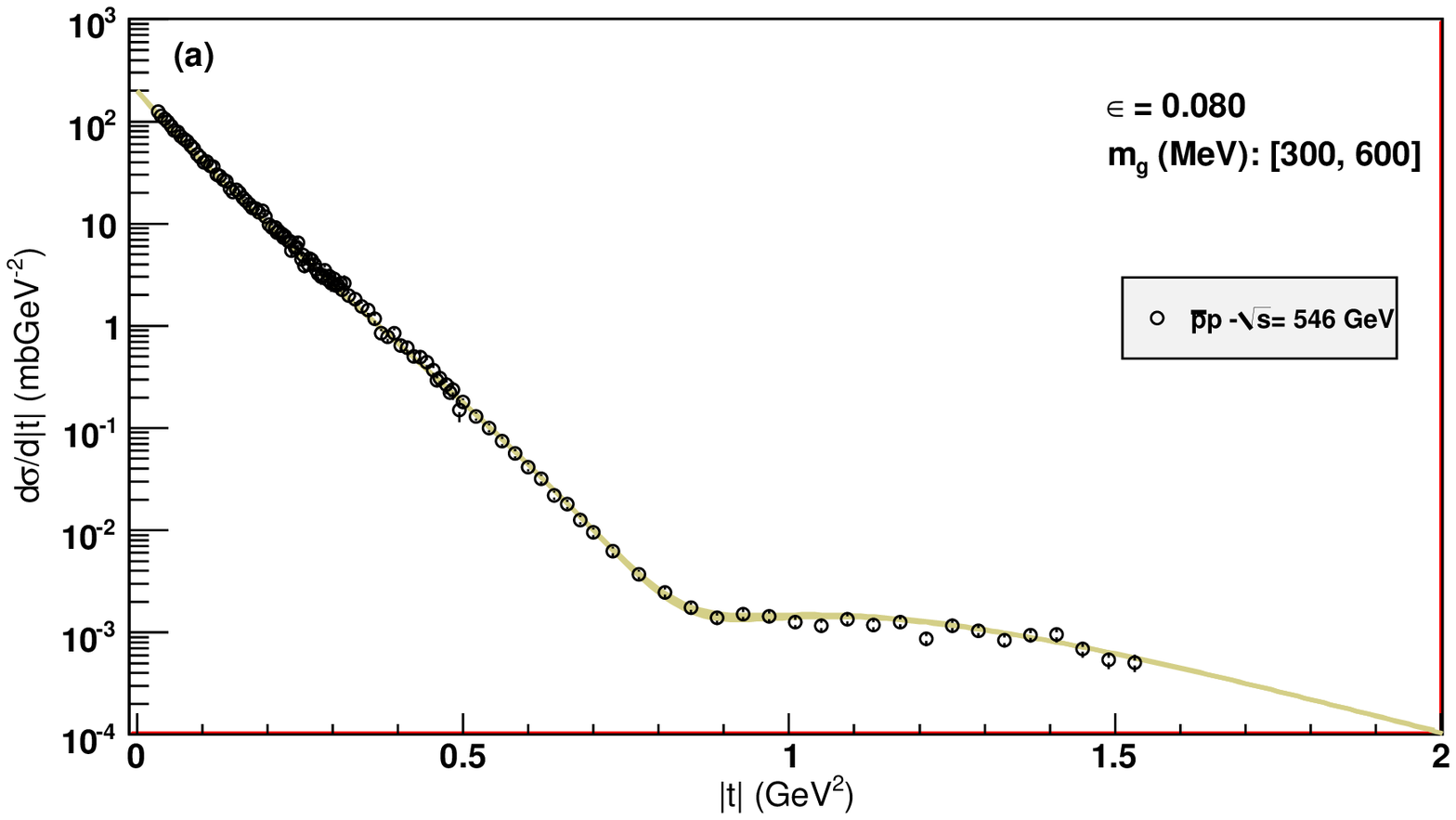,width=7.6cm,height=8cm}
\epsfig{file=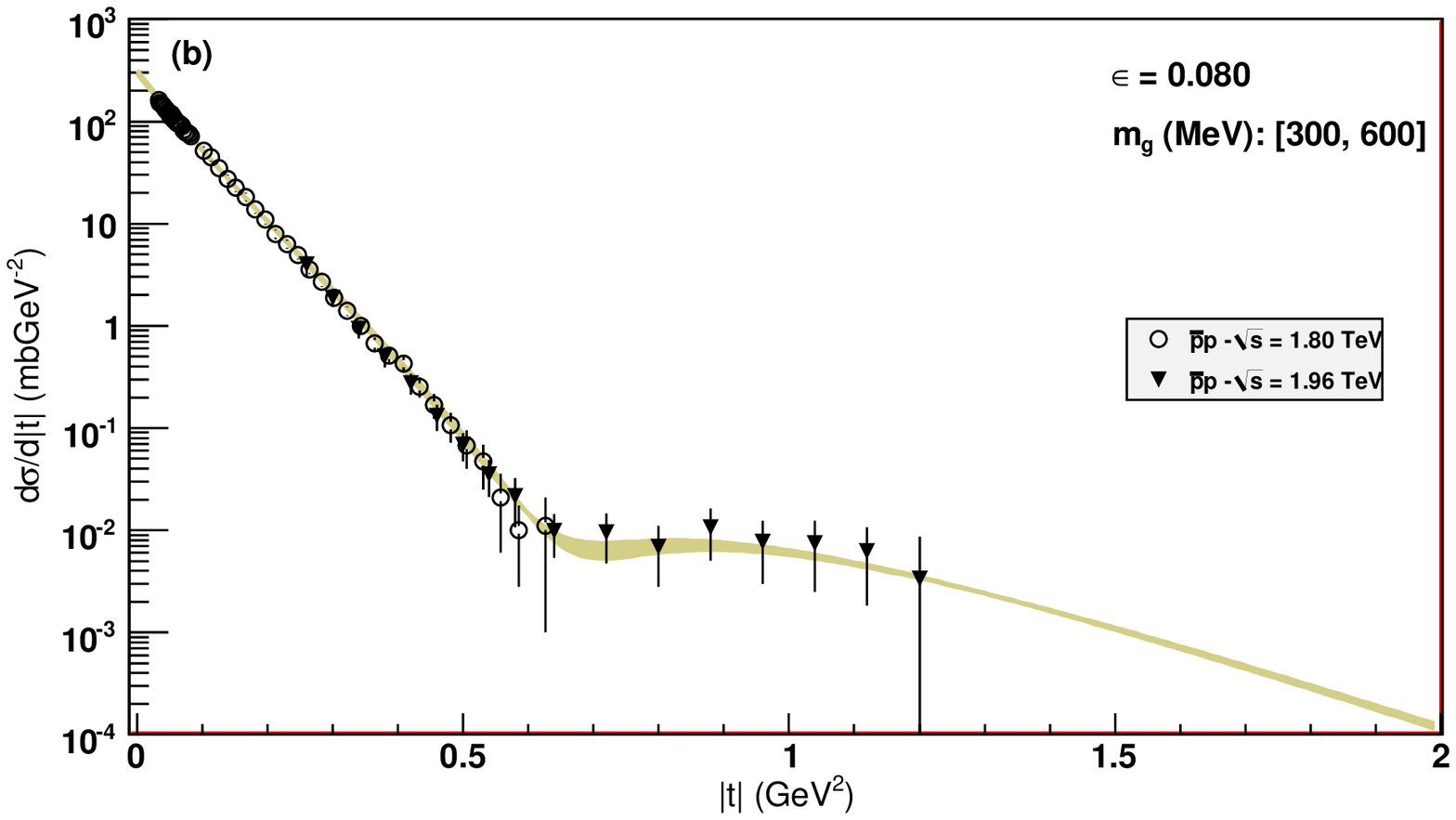,width=7.6cm,height=8cm}
\caption{Differential cross sections at (a) 546 GeV and (b) 1.8 TeV (together with the 1.96 TeV data)
for fixed $\epsilon$ = 0.080, upper bound for $m_g$ = 600 MeV
and lower bound for  $m_g$ = 300 MeV.}
\label{f3}
\end{figure}
\begin{figure}[pb]
\centering
\epsfig{file=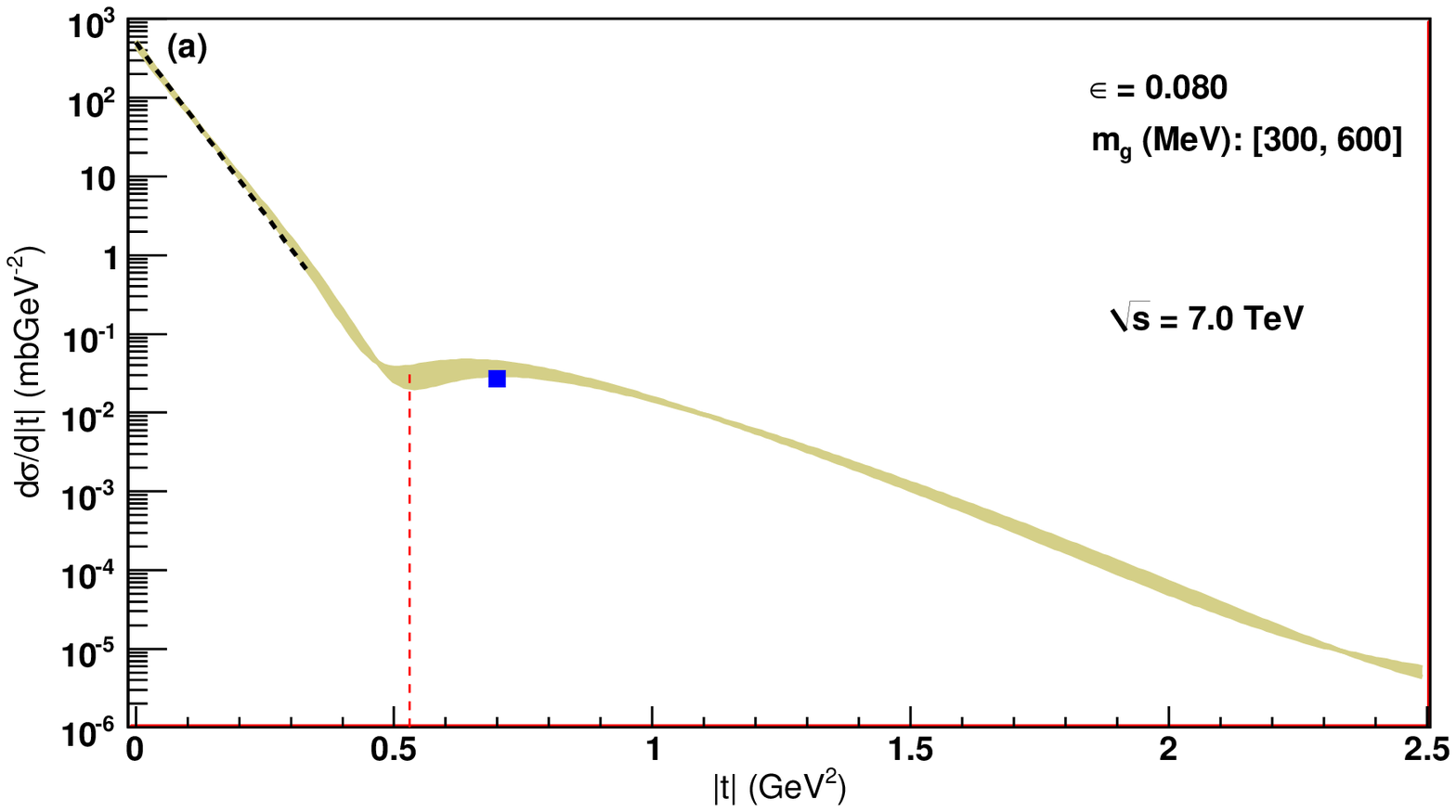,width=7.6cm,height=8cm}
\epsfig{file=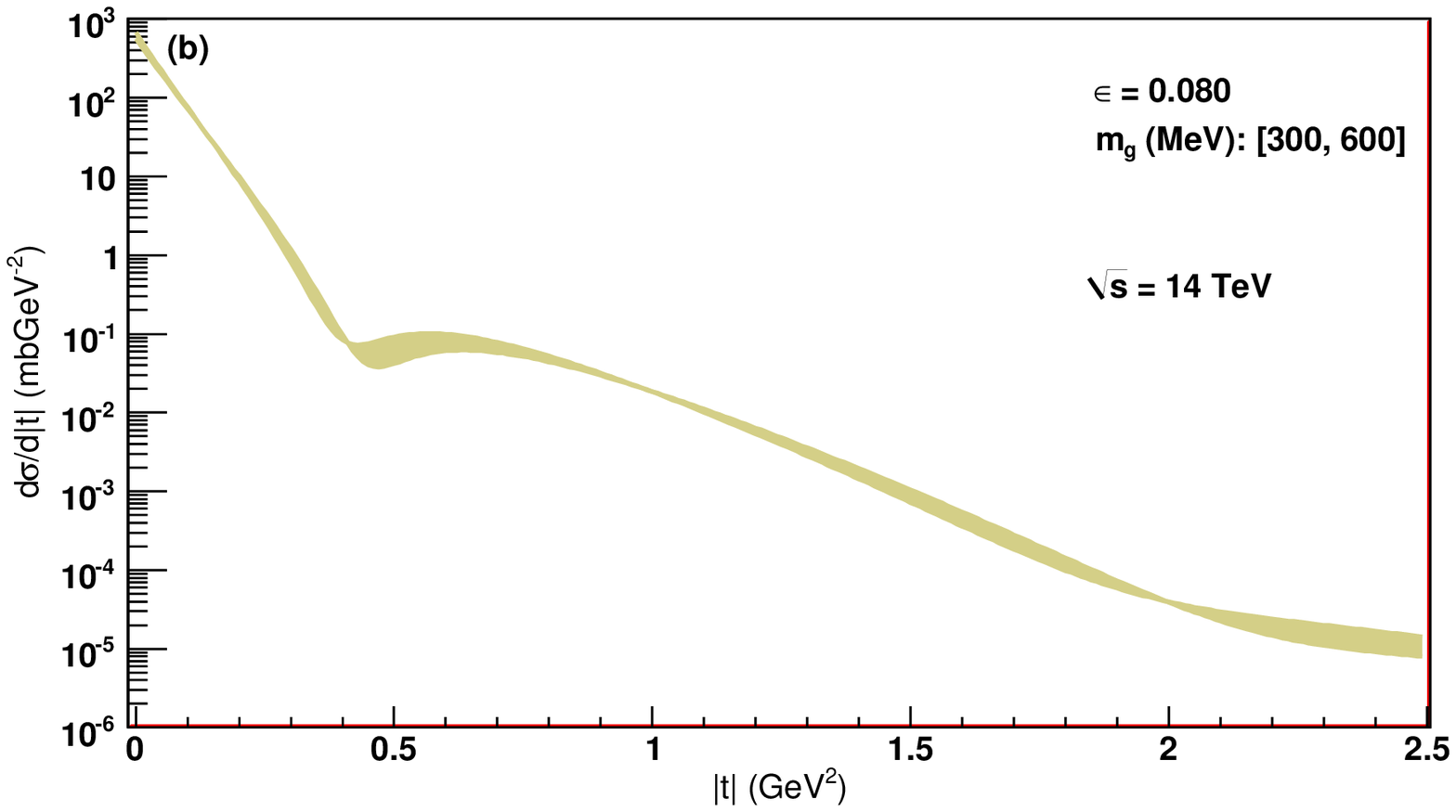,width=7.6cm,height=8cm}
\caption{Predictions for the differential cross section at 7 TeV (a) and 14 TeV (b) for 
fixed $\epsilon$ = 0.080, upper bound for $m_g$ = 600 MeV
and lower bound for  $m_g$ = 300 MeV. At 7 TeV the dotted segment indicate the optical point
and slope at the diffraction peak,
the dashed vertical segment the dip position
and the black square the differential cross section at $q^2$ = 0.7 GeV$^2$, as measured by TOTEM Collaboration
(see Table V).}
\label{f4}
\end{figure}

Let us first consider the case $\epsilon$ = 0.080 and  $m_g$ (MeV): [300,$\,$ 600]. 
The values of the fit parameters for $m_g$ = 300, 400, 500 and 600 MeV are shown in Table II.
With the extrema values $m_g$ = 300 MeV and  $m_g$ = 600 MeV for each 
evaluated quantity we plot the upper and the lower
curves delimiting the region of uncertainty represented by a band, in the figures that follows.
The results for $\sigma_{tot}(s)$, $\rho(s)$ are shown in Fig. 2 and those for the
$\bar{p}p$ differential cross sections at 546 GeV and 1.8 TeV in Fig. 3. 

For further discussion, we have included in these figures some data that did not
take part in the data reductions. In the plot of $\sigma_{tot}$ (Fig. 2.a) we have included the recent 
result
by the TOTEM Collaboration at 7 TeV \cite{totem2} and in the differential cross section at
1.80 TeV (Fig. 3.b) the data at 1.96 TeV \cite{d0}.

The predictions for the $pp$ differential cross section data at 7 TeV and 14 TeV are displayed in Fig. 4.
In the former case (7 TeV) the straight line segments and the point 
indicate recent experimental results obtained by the TOTEM Collaboration  \cite{totem1, totem2} (see Table V):
the optical point with the slope at the diffraction peak (dotted segment), the dip position
(dashed vertical segment) and the value of the differential cross section at $q^2$ =
0.7 GeV$^2$ (black square).

\begin{table}[ht]
\caption{Values of the fit parameters for $m_{g} = 400$ MeV. $C_o$, $C_{qq}$, $C_{qg}$, $C'_{qg}$ and $C_{gg}$ are dimensionless 
and  $\mu_{o}$, $\mu_{qq}$, $\mu_{qg}$, $\mu_{gg}$ are in GeV.}
\begin{center}
\begin{tabular}{cccc}
\hline
$\epsilon$: & 0.080 & 0.085 & 0.090\\
\hline
$C_{o}$                    & 3.03$\pm$0.40    & 3.10$\pm$0.48      & 3.11$\pm$0.42\\
$C{qq}$                        & 10.7$\pm$1.4     & 10.5$\pm$1.2       & 10.2$\pm$1.1\\
$C_{qg}$($\times$10$^{-1}$)    & 8.74$\pm$0.59    & 8.63$\pm$0.47      & 8.66$\pm$0.41\\
$C'_{qg}$($\times$10$^{-2}$)    & 4.51$\pm$0.62    & 4.68$\pm$0.50      & 4.69$\pm$0.45\\
$C_{gg}$($\times$10$^{-3}$)    & 3.79$\pm$0.17    & 3.62$\pm$0.14      & 3.49$\pm$0.12\\
$\mu_{o}$                  & 0.41$\pm$0.17    & 0.44$\pm$0.18      & 4.45$\pm$0.16\\
$\mu_{gg}$                 & 0.651$\pm$0.066  & 0.6500$\pm$0.0065  & 0.6496$\pm$0.0063\\
$\mu_{qq}$                 & 1.32$\pm$0.16    & 1.30$\pm$0.14      & 1.27$\pm$0.16\\
$\mu_{qg}$                 & 0.838$\pm$0.044  & 0.8367$\pm$0.0040  & 0.8367$\pm$0.0037\\
\hline
\end{tabular}
\end{center} 
\end{table}
\begin{figure}[pb]
\centering
\epsfig{file=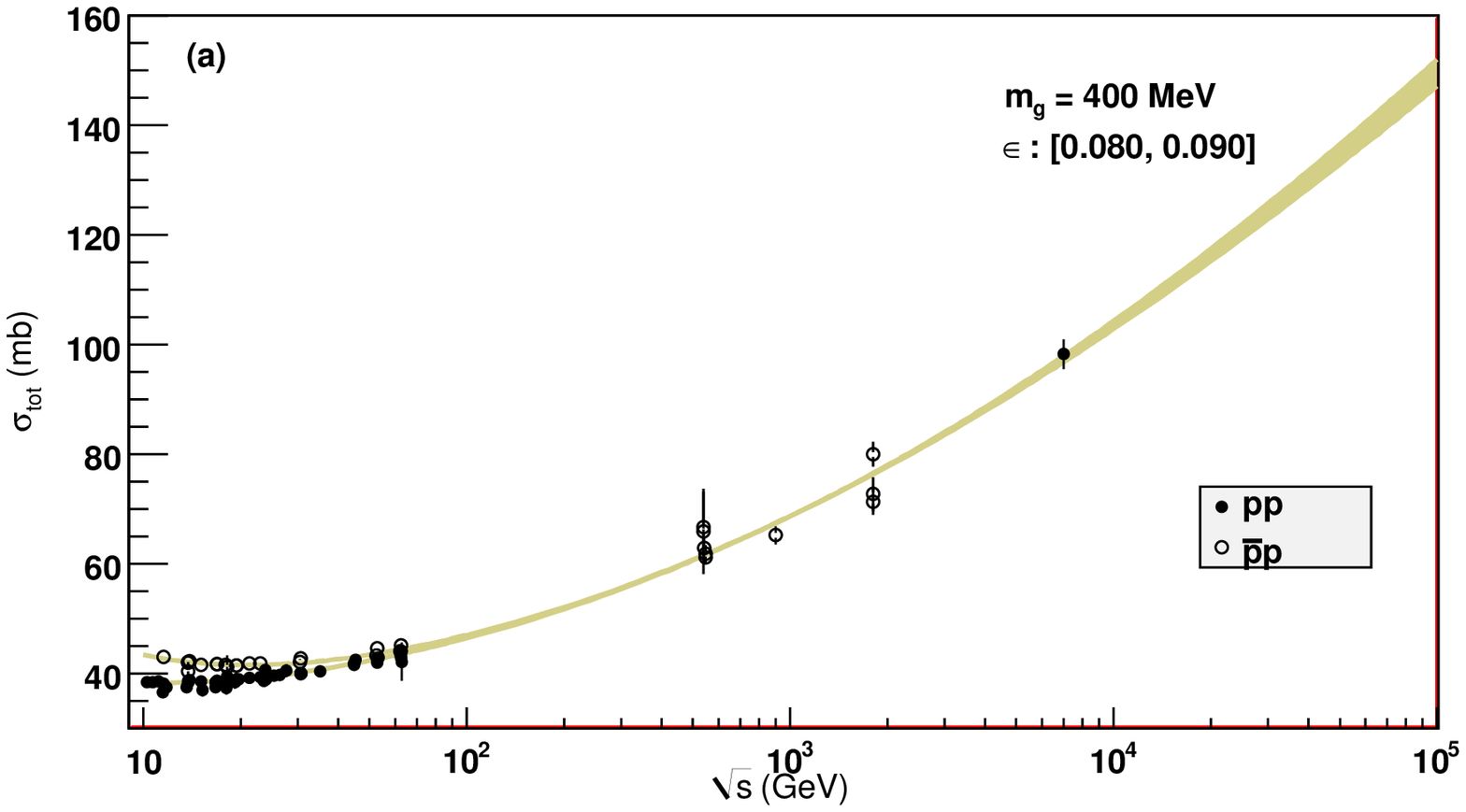,width=7.6cm,height=8cm}
\epsfig{file=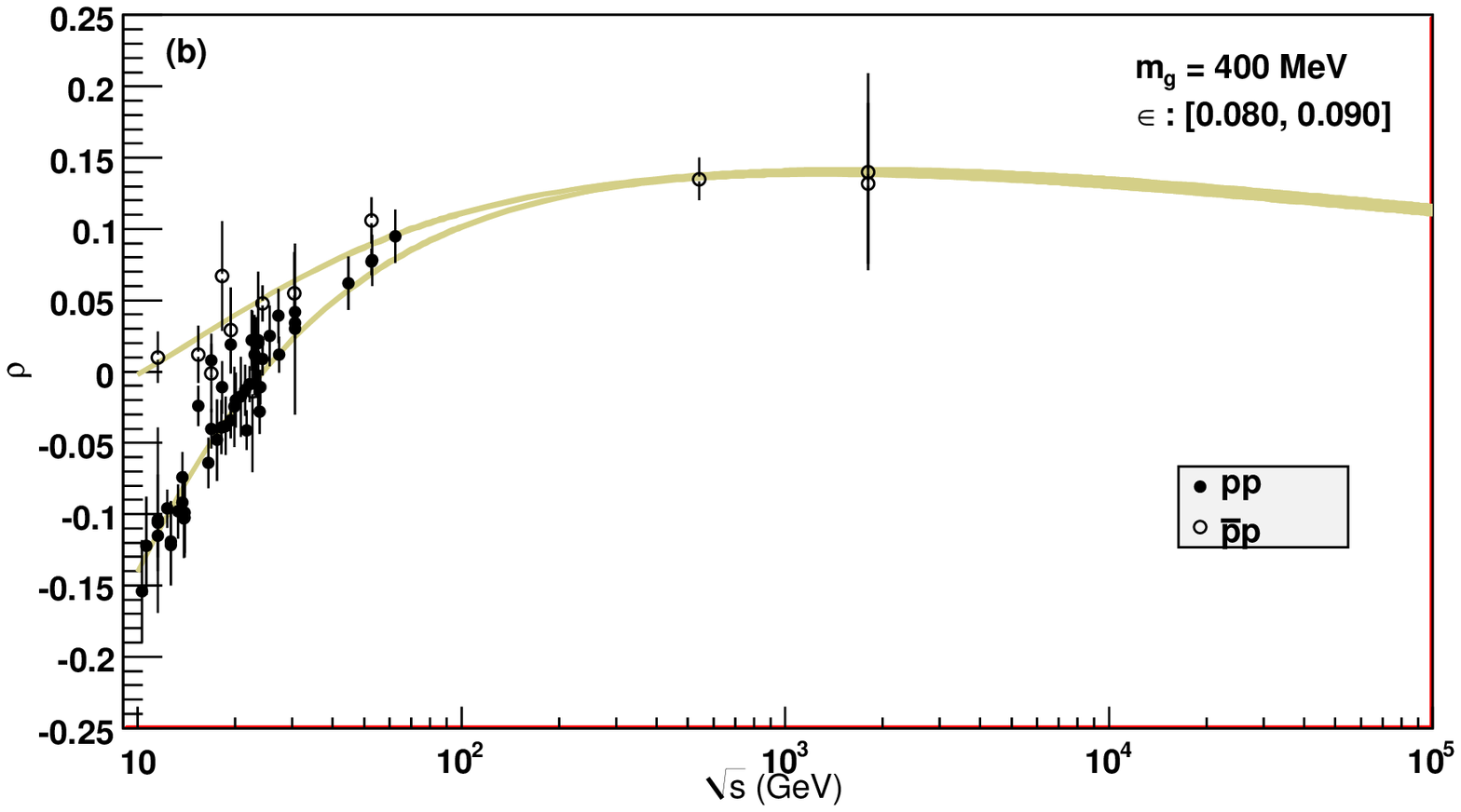,width=7.6cm,height=8cm}
\caption{Total cross section and $\rho$ parameter for fixed $m_g$ = 400 MeV, upper bound for $\epsilon$ = 0.090
and lower bound for  $\epsilon$ = 0.080.}
\label{f5}
\end{figure}
\begin{figure}[pb]
\centering
\epsfig{file=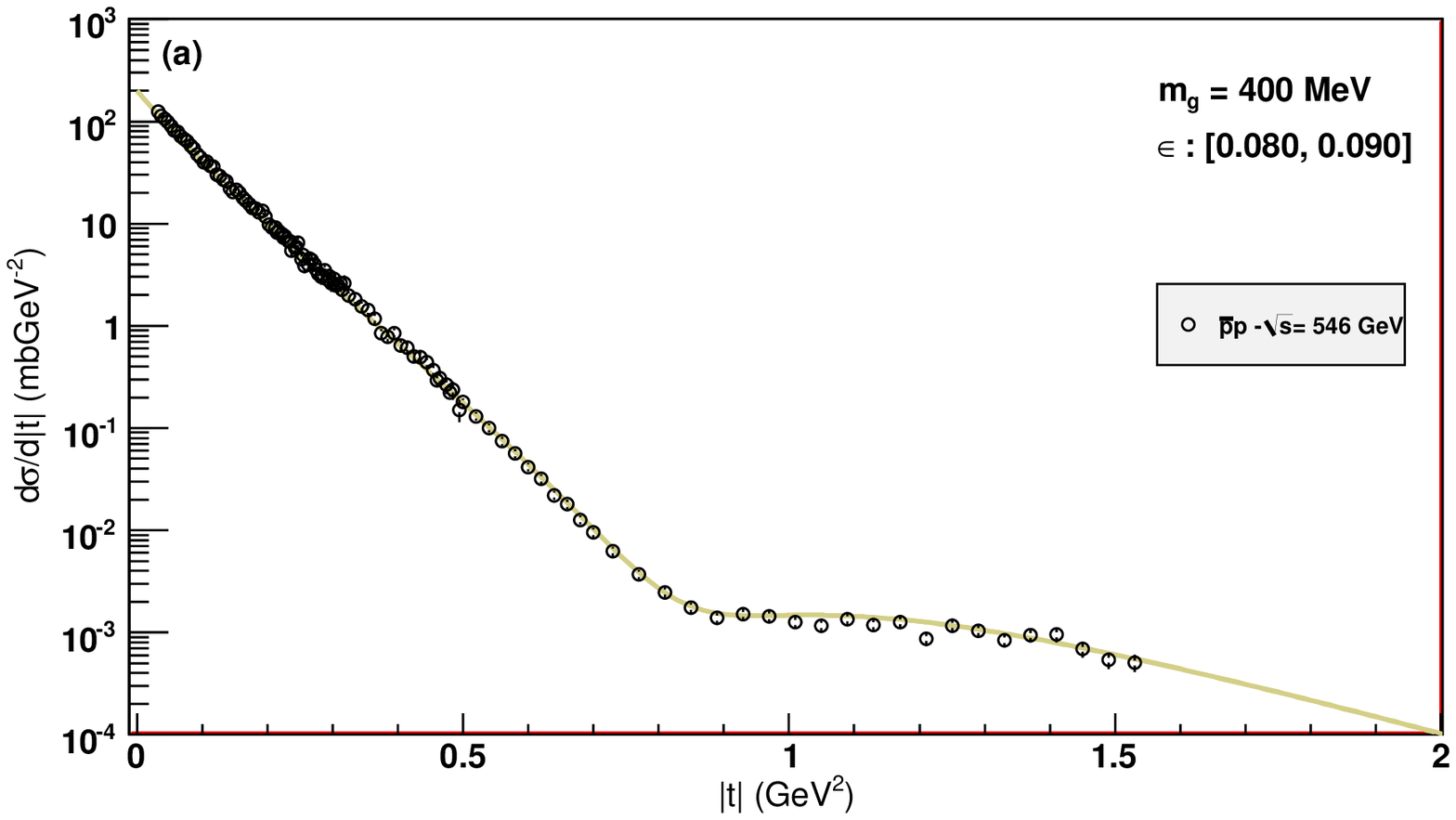,width=7.6cm,height=8cm}
\epsfig{file=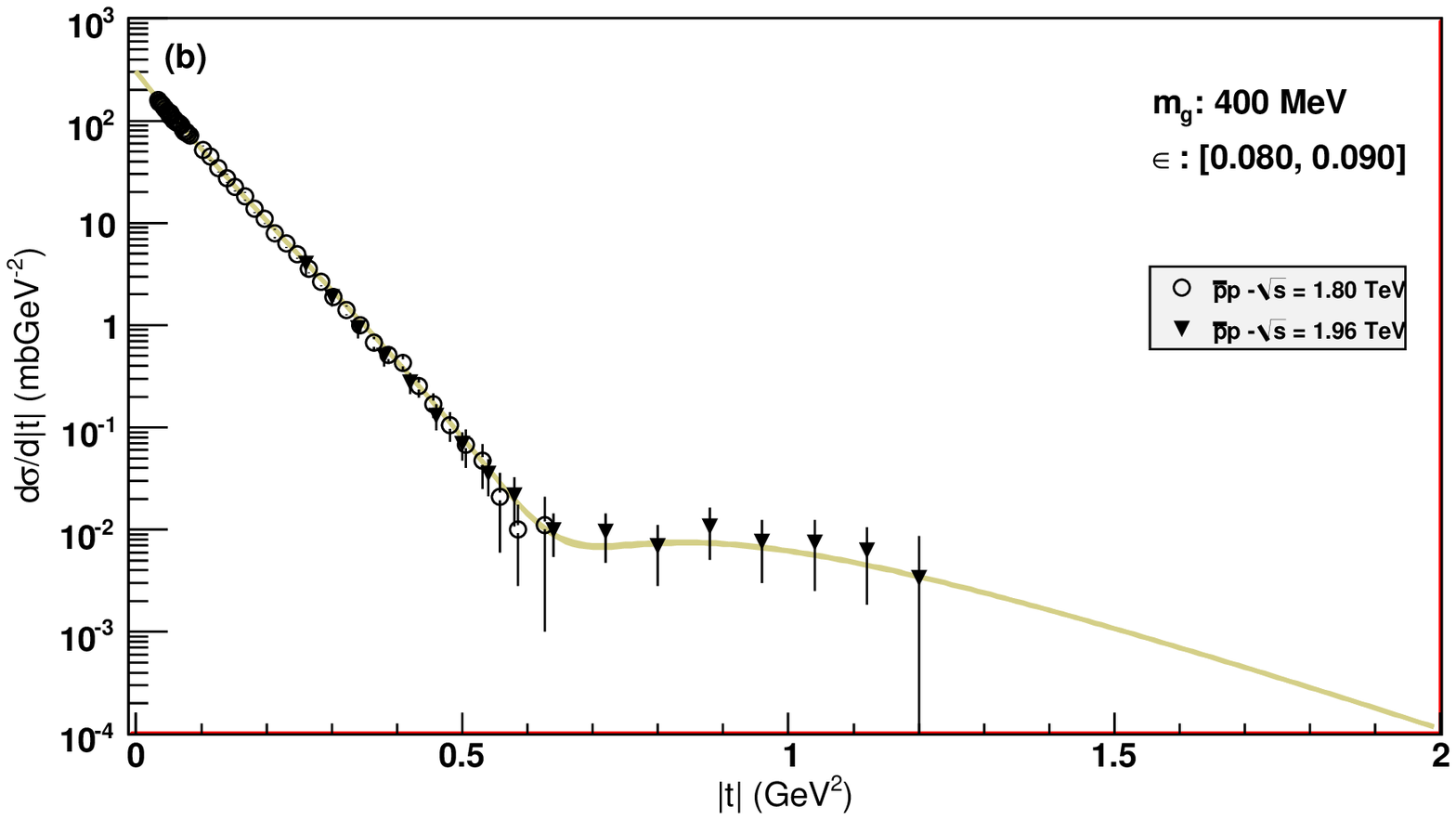,width=7.6cm,height=8cm}
\caption{Differential cross sections at (a) 546 GeV and (b) 1.8 TeV (together with the data
at 1.96 TeV) for fixed $m_g$ = 400 MeV, upper bound for $\epsilon$ = 0.090
and lower bound for  $\epsilon$ = 0.080.}
\label{f6}
\end{figure}
\begin{figure}[pb]
\centering
\epsfig{file=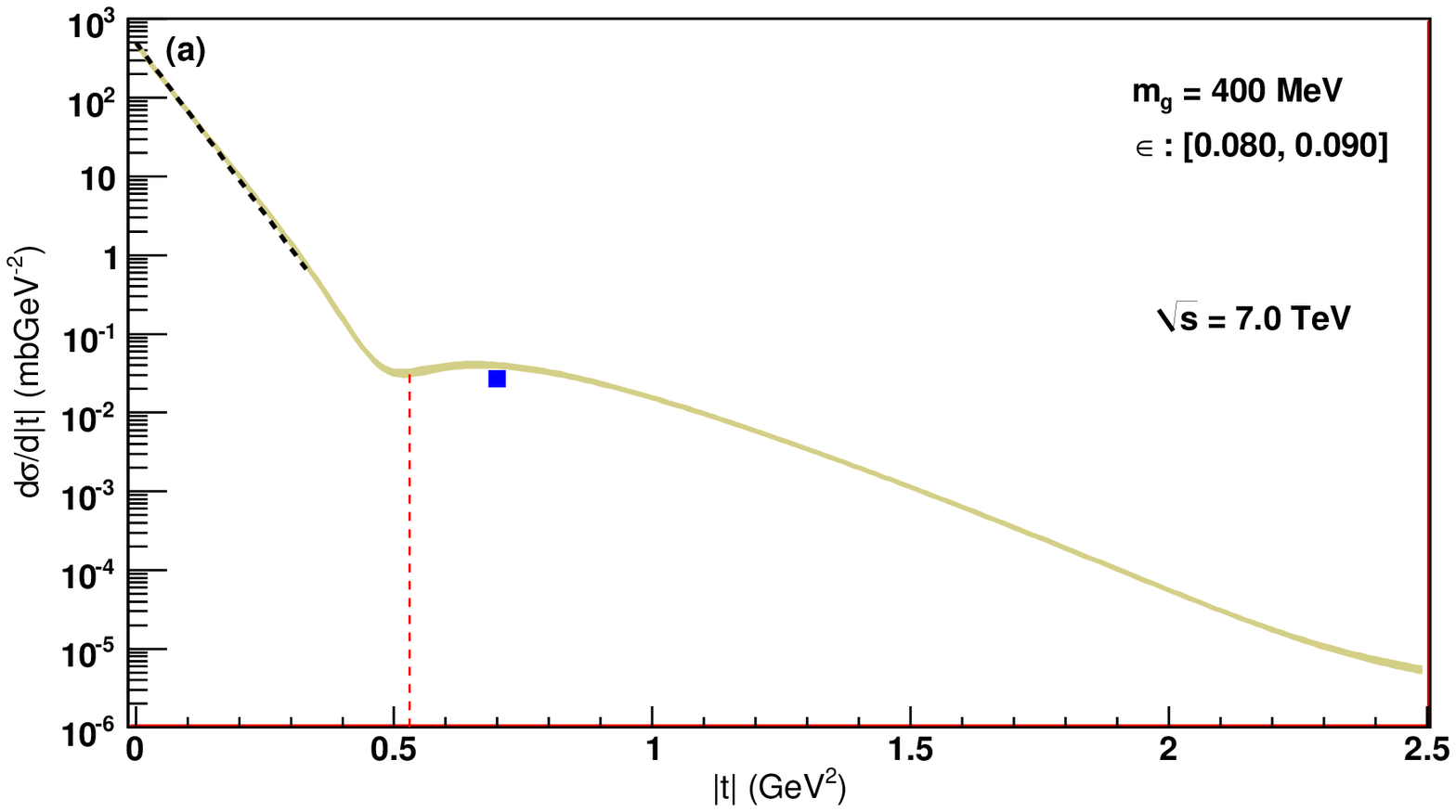,width=7.6cm,height=8cm}
\epsfig{file=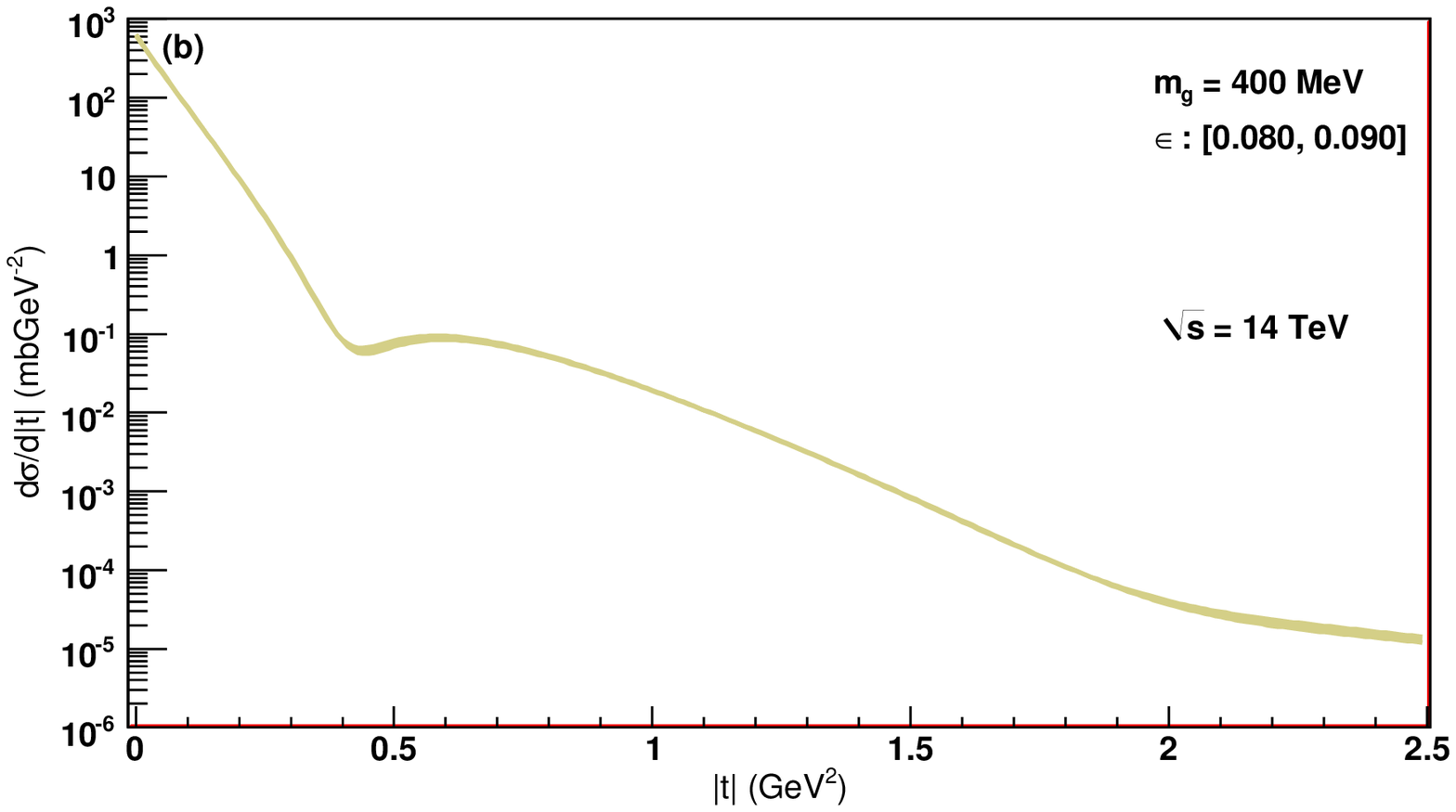,width=7.6cm,height=8cm}
\caption{Predictions for the differential cross section at 7 TeV 
(a) and 14 TeV (b) for fixed $m_g$ = 400 MeV, upper bound for $\epsilon$ = 0.090
and lower bound for  $\epsilon$ = 0.080. Same legend as in Fig. 4.}
\label{f7}
\end{figure}

For the case $m_{g} = 400$ MeV and $\epsilon$: [0.080, $\,$ 0.090],
the values of the fit parameters for $\epsilon$= 0.080, 0.085 and 0.090 are displayed in Table III.
Here, with the extrema values $\epsilon$ = 0.080 and  $\epsilon$ = 0.090 we plot the upper and the lower
curves delimiting the region of uncertainty represented by a band in the figures that follows.
The corresponding results for $\sigma_{tot}(s)$, $\rho(s)$ and the differential cross sections
are shown in Figs. 5, 6 and 7.

\subsection{Discussion}

The main point in this section has been to propose estimations of the uncertainties in
the evaluated quantities associated with relevant intervals for the fundamental
parameters $\epsilon$ and $m_{g}$. As commented before we did not address the
uncertainties associated with both intervals simultaneously but only an individual interval
for a fixed parameter.
Although that may represent an underestimation of the effective uncertainty it certainly gives more
information on the reliability of the results than what could be obtained by fixing
both parameters (or 6 as in  \cite{block1}).
In addition, in practice one band is practically contained in the other, so that the larger bands represent a
reasonable result for our purposes.

The results presented in Tables I - III, figures 2, 3, 4 and 5, 6, 7 lead to some
\textit{qualitative} conclusions:

\begin{itemize}
\item[i.]
In addition to the good quality of the statistical results in terms of $\chi^2$/DOF (Table I), 
Figs. 2, 3 and 4 show that the visual description of all reduced data is quite good.
In this energy region (up to 1.8 TeV) the inferred error bands are not significant, which means that for any fixed values of 
$\epsilon$ and $m_g$ considered in the analysis, all data are quite well described. The correct
reproduction of the high-energy differential cross section data up to
$q^2 \sim$ 1.5 GeV$^2$ (Figs. 3 and 6) seems to be a remarkable result in the
context of QCD-inspired models. 

\item[ii.]
At higher energies (above those considered in the data reductions)
the effects of the relevant intervals are significant
since the general trend is an increase in the estimated error band above 1.8 TeV
(Figs. 2 and 5).

\item[iii.]
The error bands for fixed $m_g$ (400 MeV) are narrower than those for fixed $\epsilon$ 
(0.080) and the former are, in general,  contained in the latter. That is a consequence of the difference
in the relative intervals involved, $\sim$ 6 \% for $\epsilon$ and $\sim$ 30 \% for $m_g$,
as well as the role played by each parameter in the formalism and data reductions.

\item[iv.]
Table II and III illustrate the correlations among $\epsilon$, $m_g$ and the fit
parameters. We see that this correlation is stronger in the case of $\epsilon$ fixed
than in the case of $m_g$ fixed, an effect also associated with the difference in the relative
intervals. Moreover, note that in both cases the correlations are stronger among the parameters associated
with the normalization constants
(the $C's$) than among those coming from the form factors (the $\mu's$).

\item[v.]
Taking into account the error bands, 
the total cross section result by the TOTEM Collaboration at 7 TeV \cite{totem2}
is quite well described in both cases (Fig. 2.a and Fig. 5.a) and a reasonable consistency can
be observed in
the differential cross section up to the dip position (Fig. 4.a and Fig. 7.a).

\item[vi.]
For the differential cross section at 14 TeV a change of curvature is predicted around
$q^2$ = 2 GeV$^2$ in both cases (Fig. 4.b and Fig. 7.b).
\end{itemize}

Let us now present and discuss some \textit{quantitative} results and predictions with the 12 variants considered,
as well as the inferred upper and lower bounds for the evaluated quantities. For each one of the 12 variants
the results for $\sigma_{tot}$, $\rho$, $\sigma_{in}$ and $\sigma_{el}/\sigma_{tot}$ at 7, 14 and 57 TeV
are displayed in Table IV. As commented before, with the extrema values of $\epsilon$ or $m_g$
in each relevant interval, we evaluate the upper and lower bounds for all evaluated quantities.
Table IV shows that in fact, for the values of $\epsilon$ and $m_g$ inside the relevant intervals
the numerical values of the evaluated quantities lies inside the upper and lower bounds.
Therefore, in what follows we shall give our predictions in terms of the band corresponding
to these upper and lower bounds in each case ($\epsilon$ or $m_g$ fixed). Since these bands have
the same meaning as the relevant intervals
(equally likely  values),
we shall use the same notation, namely
the bounds inside brackets and separated by a comma: [lower-bound, upper-bound].

\begin{table}[ht]
\caption{Predictions for $pp$ integrated cross sections and $\rho$ parameter at the LHC 
and Auger energies for each tested value of $m_{g}$ and $\epsilon$. For each physical
quantity the extrema values define the uncertainty band. The cross sections
are in mb.}
\begin{center}
\begin{tabular}{c|ccccc|cccc}
\hline\hline
physical     &  $\epsilon$ = 0.080&    &    &   &             &  $m_g$ = 400 MeV  &   & \\
quantity     & $m_g$ (MeV) : & 300 &400 &500 &600& $\epsilon$: & 0.080 & 0.085 & 0.090\\
\hline
$\sigma_{in}$ (7 TeV) &  & 70.0 & 72.3& 70.6 & 74.6 &   &72.3 & 72.6& 73.0\\
$\sigma_{in}$ (14 TeV)&  & 76.7&  79.8& 77.5 & 82.9 &   &79.8 & 80.3& 81.0\\
$\sigma_{in}$ (57 TeV) & & 91.4&  96.1& 92.8 & 100.9 &  &96.1 & 97.2& 98.4\\
\hline
$\sigma_{tot}$ (7 TeV) & & 93.2               & 96.9 & 94.0 & 100.2&             & 96.9&97.4 & 98.0\\
$\sigma_{tot}$ (14 TeV)& & 103.9              & 108.8 & 104.9 & 113.4&           &108.8 &109.6 & 110.6\\
$\sigma_{tot}$ (57 TeV)& & 127.8              & 135.6 & 129.5 & 143.1 &          &135.6 &137.3 & 139.4\\
\hline
$\sigma_{el}/\sigma_{tot}$(7 TeV)&  & 0.2489  & 0.2539& 0.2489& 0.2554&          & 0.2539 & 0.2546 & 0.2551 \\
$\sigma_{el}/\sigma_{tot}$(14 TeV)& & 0.2618  & 0.2665& 0.2612& 0.2690&          & 0.2665 & 0.2673 & 0.2676 \\
$\sigma_{el}/\sigma_{tot}$(57 TeV & & 0.2848  & 0.2834& 0.2834& 0.2949&          &  0.2834 & 0.2921 & 0.2941 \\
\hline 
$\rho(7\ TeV)$ & & 0.1225                     & 0.1321   & 0.1245  & 0.1409 &    & 0.1321 & 0.1346& 0.1376\\
$\rho(14\ TeV)$& & 0.1189                     & 0.1272   & 0.1208  & 0.1349 &    & 0.1272 & 0.1299& 0.1330 \\
\hline\hline
\end{tabular} 
\end{center} 
\end{table}

Let us consider first the $pp$ elastic scattering at 7 TeV with focus on the
published results by the TOTEM Collaboration \cite{totem1, totem2}. Our predictions
(bands) for several physical quantities are displayed in Table V, together with the
corresponding results by the TOTEM Collaboration. From that Table and within the error bands
we are led to the conclusions that follows.

$\bullet$ For $\epsilon$ = 0.080 the predictions are: (a) consistent with all cross sections,
optical point and dip position; (b) barely consistent with the forward slope;
(c) not consistent with the slope near the dip position, the differential
cross section at $q^2$ = 0.7 GeV$^2$ and the exponent $n$ in the power
law  at large momentum transfer.

$\bullet$ For $m_g$ = 400 MeV the predictions are: (a) consistent with all the cross
sections and the optical point;
(b) barely consistent with the dip position;
(c) not consistent with both slopes, the differential
cross section at $q^2$ = 0.7 GeV$^2$ and the exponent $n$.

How good or bad are these results?
In \cite{totem1} the TOTEM Collaboration displays in Table 4 the predictions
from some representative phenomenological models
compiled in \cite{kkl}. Comparison of our results with these predictions
shows that despite all the above inconsistencies, our  efficiency  in the description of the experimental
data is compatible with that presented by all quoted models. It is also interesting to note in that Table
that our wrong prediction for the exponent in the power law at large $q^2$, namely
$n \sim$ 10.5, is in plenty agreement with the prediction by Block \textit{et al}.:
$n$ = 10.4. This agreement may be related with the choice of four dipole form factors,
present in both models. Moreover, the first TOTEM data on differential elastic cross section at 7 TeV has been measured in specialized runs
of LHC and in the momentum transfer range of $0.36<-t<2.5$ GeV$^{2}$. Notice that none of the representative models
for elastic scattering can fully reproduce its overall features; in special, the dip position, dip depth and large-t behaviour \cite{totem1}.
Particularly in the range $1.5<-t<2.0$ GeV$^{2}$ the data presented a power law pattern proportional to $\sim |t|^{-8}$, which
is in poorly agreement with general model predictions. Interestingly enough, this specific power law behaviour was first envisaged 
by Donnachie and Landshoff \cite{dl} in a model which accounts for the contribution of three gluon exchange in elastic scattering 
at high energies. Yet, in the QCD context, the constituent interchange model by Lepage and Brodsky \cite{bl} predicts a different power 
law behaviour, namely $\sim |t|^{-10}$. The DGM model results at the large-$t$ region are consistent with the latter.
We shall return to this point in our conclusions.

At other energies (14 and 57 TeV) the bands in our predictions also correspond to the extrema
values for each physical quantity, as given in Table IV, for the cases $\epsilon$ = 0.080
and $m_g$ = 400 MeV. For example, at 57 TeV we predict 
$\sigma_{tot}$ (mb): [127.8, $\,$ 143.1] for $\epsilon$ = 0.080 and 
$\sigma_{tot}$ (mb): [127.8, $\,$ 139.4] for $m_g$ = 400 MeV. We note that these results are
consistent with recent estimation of this quantity through an analytical parametrization,
$\sigma_{tot}$ = 133.4 $\pm$ 1.6 mb \cite{bh}.

\begin{table}[ht]
\caption{TOTEM results at 7 TeV with systematic and statistical
error added in quadrature and the uncertainty band predictions.}
\begin{center}
\begin{tabular}{|c|c|c|c|}
\hline\hline
Physical   & TOTEM                        & $\epsilon$ = 0.080           & $m_g$ = 400 MeV \\
quantity   & results  & $m_g$ (MeV): [300, 600] & $\epsilon$: [0.080,  0.090] \\
\hline
B($0.36 \leq t \leq$0.47 GeV$^{2}$) (GeV$^{-2}$) \cite{totem1}& 23.60  $\pm$ 0.64 & [19.8,  22.4] & [20.8, 21.6]\\
\hline
$|t_{dip}|$ (GeV$^{2}$) \cite{totem1} & 0.53 $\pm$ 0.01 & [0.51, 0.54] & [0.51, 0.52] \\
\hline
$n$ in $|t|^{-n}$ ($1.5 \leq t \leq$2.0 GeV$^{2}$) \cite{totem1} & 7.80 $\pm$ 0.32 & [10.1, 10.6] & [10.5, 10.5]\\
\hline
$\frac{d\sigma}{d|t|}(|t| = 0.7)$ (mbGeV$^{-2}$) \cite{totem1} & 2.70$^{+0.71}_{-0.58}\times$10$^{-2}$ &
[3.0, 4.3]$\times 10^{-2}$ & [3.8, 4.1]$\times 10^{-2}$\\
\hline
$\sigma_{el}$ (mb) \cite{totem2}& 24.8 $\pm$ 1.2 & [23.2, 25.6] & [24.6,  25.0]\\
\hline
$\sigma_{in}$ (mb) \cite{totem2}& 73.5$^{+1.9}_{-1.4}$  & [70.0, 74.6] & [72.3, 73.0] \\
\hline
$\sigma_{tot}$ (mb) \cite{totem2}& 98.3 $\pm$ 2.8 & [93.2, 100.2] & [96.9, 98.0] \\
\hline
B($0.02 \leq t \leq$0.33 GeV$^{2}$) (GeV$^{-2}$) \cite{totem2}& 20.10  $\pm$ 0.36 & [18.6, 19.8] & [19.3, 19.4] \\
\hline
$\frac{d\sigma}{d|t|}\arrowvert_{t=0}$ (mbGeV$^{-2}$) \cite{totem2}& 504 $\pm$ 27 & [451, 523] & [488, 500] \\
\hline\hline
\end{tabular}
\end{center}
\end{table}

At last, let us discuss our two proposed independent solutions for the inference of uncertainties.
Guided by the TOTEM results at 7 TeV, Table V suggests that the case of fixed $\epsilon$ = 0.080 presents
better consistence with the experimental data than the case of fixed $m_g$ = 400 MeV. Since we shall not treat
the simultaneous effects of both parameters, we may conclude that by fixing $\epsilon$ at
the above value we arrive at a best solution. Indeed this conclusion is corroborated by the following facts:

$\bullet$ The bands for fixed $m_g$ lies, in general, inside those for fixed $\epsilon$
and therefore the former may be seen as a particular case of the latter.

$\bullet$ The fixed value $\epsilon$ = 0.080 for the soft Pomeron intercept is consistent with all phenomenological
analyses of the elastic hadron scattering, in particular with those treating
constraint and extrema bounds \cite{lm,lmm}.

$\bullet$  The interval $m_g$ (MeV): [300,$\,$ 600] is consistent with all phenomenological and theoretical results on the
dynamical gluon mass.

Therefore, we understand that our approach, as it has been presented here, has an optimum solution
with only one fixed parameter, $\epsilon$ = 0.080 and a relevant physical interval for the dynamical
gluon mass scale, $m_g$ (MeV): [300,$\,$ 600]. In this case, within the inferred error bands
the predictions present reasonable consistency with the TOTEM results at 7 TeV, at least
at the same level of some representative phenomenological models. 

\section{Conclusions and Final remarks}

Further developments on a dynamical gluon mass approach to elastic $pp$ and $\bar{p}p$ scattering
have been presented and discussed. The main conceptual ingredient, introduced in \cite{luna01,luna02}, concerns
the dynamical gluon mass scale as a natural regulator for the IR divergences associated with the
gluon-gluon cross section, leading to the identification of explicit nonperturbative contributions to
high-energy elastic hadron scattering.
In this context a consistent physical meaning is given to two \textit{ad hoc} parameters, typical of
mini-jet or QCD-inspired models: the infrared mass scale and the effective value of the running coupling
constant.

In this paper the main novel results concern the account of the energy dependencies in the dynamical
gluon mass, Eq. (8), the detailed investigation on the influence
in the predictions from relevant physical intervals for the dynamical gluon mass scale
and the soft Pomeron intercept, as well as the introduction of a method to estimate
the corresponding bands of uncertainty in all evaluated quantities.
Fits to an improved data ensemble on $pp$ and $\bar{p}p$ elastic scattering up to 1.8 TeV have led to quite good
descriptions of all reduced data, including the high-energy differential cross section
up to $q^2 \sim$ 1.5 GeV$^2$. With the inferred uncertainty bands the experimental data recently obtained by
the TOTEM Collaboration are reasonably well described, except the region beyond the dip position.
Comparison with the predictions from some representative phenomenological
models, presented and quoted in \cite{totem1}, shows that our results are at the same level
of consistency with the TOTEM data,
with the advantage in our case of explicit connections with nonperturbative QCD.
As commented in Appendix A, our data reductions disfavor a soft solution for the
dynamical gluon mass.

The uncertainty in the evaluated quantities, associated with relevant intervals for the fundamental
parameters $\epsilon$ and $m_{g}$, seems to us a fundamental information
for reliable phenomenological predictions.
Although we did not address the question associated
with both intervals simultaneously, we understand that our proposed estimation of the bounds
gives more
information on the reliability of the results than what could be obtained by fixing
both parameters.
Based on the results and discussion displayed in Subsection 3.3, we propose as our best present solution
the case of fixed
$\epsilon$ = 0.080 and interval $m_g$ (MeV): [300,$\,$ 600]. With that we have only one fixed parameter,
with well founded physical justification.

The critical point raised in respect to fixing parameters have been directed to a class of QCD inspired models. However
the criticism certainly apply to any phenomenological model with fixed parameters whose numerical values
do not have an explicit physical justification and its consequences in the evaluated quantities
are not investigated or even discussed.

In the phenomenological context, model developments are in general more important than any particular formulation
and, in our case, at least two fundamental aspects demand further investigation.
The first one concerns the phenomenological choice for the gluon distribution function, Eq. (18),
which introduces the Pomeron intercept as a fundamental parameter. Despite its efficiency
in the analysis we have developed, it may be important to study the effects of
different parton density functions, along the lines that have been discussed by Achilli \textit{et al}. \cite{achilli}.
In this class of QCD-based models, the attenuation in the rise of the total cross sections comes from soft gluon
emission from colliding partons \cite{giulia01,giulia02,giulia03}. These emissions affect the matter distribution and the resulting overlap function can be calculated from 
the soft gluon transverse-momentum resummed distribution. In this $k_{T}$ resummation scheme 
the overlap function in b-space depends upon the behaviour of the coupling $\alpha_{s}(k_{T})$ as $k_{T} \to 0 $,
being the calculation only possible after the introduction of a phenomenological infrared modification for $\alpha_{s}(k_{T})$.
In the context of the DGM approach, it should be stressed that the QCD effective charge (7) is obtained from the QCD Lagrangian,  
i.e., it is derived from first principles, in a scenario where infrared effects are naturally taken into account. Hence 
this effective strong coupling could be naturally embodied in a soft $k_{T}$ resummation scheme in order to avoid infrared divergences.
A second aspect is related to the choice of form factors as simple dipole parametrizations. As commented
before, the fact that our prediction for the exponent in the power law at large momentum
transfer is near 10 (as is the case with the Aspen model),
suggest that this result
is directly related with the above choice for the form factors. Since
deviations from the dipole parametrization  have been indicated in model-independent analyses
of $pp$ elastic scattering \cite{fm, am08, cmm}, to take into account these empirical
results may give new insights in the predictions at large momentum transfer.
We are presently investigating the above two lines.

\appendix
\section*{Appendix A}

With the DGM approach described in Sect. 2, in addition to the Cornwall's solution (8),
we have also considered the possibility of a soft
behavior for the dynamical gluon mass.
A power-law running behavior for $M_{g}^{2}(\hat{s})$ was first envisaged
in \cite{ap1} and according to an OPE calculation the most probable
asymptotic behavior of the running gluon mass is proportional
to $1/\hat{s}$ \cite{ap2}. At the level of a non-linear Schwinger-Dyson equation
this asymptotic behavior is given by

\begin{eqnarray}
M_{g}^{2}(\hat{s}) = \frac{m_{g}^{4}}{\hat{s}+m_{g}^{2}} \left[ \frac{\ln
\left( \frac{\hat{s}+  m_{g}^{2}}{\Lambda^{2}} \right)}{
\ln \left( \frac{\tau m_{g}^{2}}{\Lambda^{2}} \right)}
\right]^{\gamma_{2}-1}  \, ,\nonumber
\end{eqnarray}
where
\[
\gamma_2 = \frac{4}{5}+ \frac{6 c_1}{5},
\]
$c_1 \in [0.7,1.3]$ and the parameters $\tau$ and $m_g$ are constrained to lie in the interval
[1.0, 8.0] and [300, 600 MeV], respectively \cite{ap3}.
However, with the procedure described in Subsect. 3.A, this solution leads to much worse $\chi^2$/DOF compared
to the ones related to Eq. (8), which disfavor such solution.

\section*{Acknowledgments}
Research supported by CNPq (A.A.N) and FAPESP (D.A.F and M.J.M).


\begin{thebibliography}{99}


\bibitem{totem1}
G. Antchev \textit{et al}. (TOTEM Collaboration)
 Europhys. Lett. \textbf{95} 41001 (2011) 

\bibitem{totem2}
G. Antchev \textit{et al}. (TOTEM Collaboration),
Europhys. Lett. \textbf{96} 21002 (2011) 

\bibitem{minijet}
D. Cline, F. Halzen, J. Luthe, Phys. Rev. Lett. \textbf{54}, 757 (1985);
T.K. Gaisser and F. Halzen, Phys. Rev. Lett. \textbf{54}, 1754 (1985);
P. L'Heureux, B. Margolis, P. Valin, Phys. Rev. D {\bf 32}, 1681 (1985);
G. Pancheri and Y.N. Srivastava, Phys. Lett. B \textbf{159}, 69 (1985);
G. Pancheri and Y.N. Srivastava, Phys. Lett. B \textbf{182}, 199 (1986);
L. Durand and H. Pi, Phys. Rev. Lett. {\bf 58}, 303 (1987);
A. Capella, J. Tran Thanh Van, J. Kwiecinski, Phys. Rev. Lett. {\bf 58}, 2015 (1987);
J. Dias de Deus, J. Kwiecinski, Phys. Lett. B \textbf{196}, 537 (1987);
L. Durand and H. Pi, Phys. Rev. D {\bf 38}, 78 (1988).

\bibitem{others}
F. Nemes, T. Csorgo,  arXiv:1202.2438 [hep-ph]; F. Nemes, T. Csorgo, arXiv:1204.5617 [hep-ph]; A.D. Martin, M.G. Ryskin, V. A. Khoze,
arXiv:1110.1973 [hep-ph].


\bibitem{giulia01} A. Corsetti, A. Grau, G. Pancheri, and Y. N. Srivastava, Phys. Lett. B \textbf{382}, 282 (1996).

\bibitem{giulia02} A. Grau, G. Pancheri, and Y. Srivastava, Phys.Rev. D \textbf{60}, 114020 (1999).

\bibitem{giulia03} R. M. Godbole, A. Grau, G. Pancheri, and Y. N. Srivastava, Phys. Rev. D \textbf{72}, 076001 (2005).

\bibitem{luna01} E.G.S. Luna, A.F. Martini, M.J. Menon, A. Mihara, and A.A. Natale,
in: AIP Conference Proceedings, vol. \textbf{739}, American Institute of Physics, New York, 2004, p. 572.


\bibitem{luna02} E.G.S. Luna, A.F. Martini, M.J. Menon, A. Mihara, and A.A. Natale, 
Phys. Rev. D {\bf 72}, 034019 (2005).

\bibitem{luna03} E.G.S. Luna and Natale, Phys. Rev. D {\bf 73}, 074019 (2006).

\bibitem{lttc01} F.D.R. Bonnet, et al., Phys. Rev. D 64 (2001) 034501;
A. Cucchieri, T. Mendes, A. Taurines, Phys. Rev. D 67 (2003) 091502(R);
P.O. Bowman, et al., Phys. Rev. D 70 (2004) 034509;
A. Sternbeck, E.-M. Ilgenfritz, M. Muller-Preussker, A. Schiller, Phys. Rev. D 72 (2005) 014507;
A. Sternbeck, E.-M. Ilgenfritz, M. Muller-Preussker, Phys. Rev. D 73 (2006) 014502;
Ph. Boucaud, et al., JHEP 0606 (2006) 001;
P.O. Bowman, et al., hep-lat/0703022;
I.L. Bogolubsky, E.M. Ilgenfritz, M. Muller-Preussker, A. Sternbeck, Phys. Lett 676 (2009) 69;
O. Oliveira, P. J. Silva, arXiv:0910.2897 [hep-lat];
O. Oliveira, P. J. Silva, arXiv:0911.1643 [hep-lat];
A. Cucchieri, T. Mendes, E.M.S. Santos, Phys. Rev. Lett. 103 (209) 141602;
A. Cucchieri, T. Mendes, Phys. Rev. D 81 (2010) 016005;
D. Dudal, O. Oliveira, N. Vandersickel, Phys. Rev. D 81 (2010) 074505.

\bibitem{pheno01} F. Halzen, G. Krein, A.A. Natale, Phys. Rev. D \textbf{47}, 295 (1993);
M.B. Gay Ducati, F. Halzen, A.A. Natale, Phys. Rev. D \textbf{48}, 2324 (1993);
A.C. Aguilar, A. Mihara, A.A. Natale, Phys. Rev. D \textbf{65}, 054011 (2002);
E.G.S. Luna, Phys. Lett. B \textbf{641}, 171 (2006);
E.G.S. Luna, Braz. J. Phys. \textbf{37} (2007) 84;
E.G.S. Luna, in: AIP Conference Proceedings, vol. \textbf{1296}, American Institute of Physics, New York, 2010, p. 183;
E.G.S. Luna, A.L. dos Santos, in: AIP Conference Proceedings, vol. \textbf{1296}, American Institute of Physics, New York, 2010, p. 330.

\bibitem{lnz} E.G.S. Luna, A.A. Natale and C.M. Zanetti, Int. J. Mod. Phys. A \textbf{23}, 151 (2008).

\bibitem{lns} E.G.S. Luna, A.A. Natale, and A.L. dos Santos, Phys. Lett. B {\bf 698}, 52 (2011).

\bibitem{lishep}
D.A. Fagundes, E.G.S. Luna, M.J. Menon, A.A. Natale, 
Testing parameters in an eikonalized dynamical gluon mass model,
arXiv:1108.1206 [hep-ph].

\bibitem{pred}
V. Barone and E. Predazzi, {\it High-Energy Particle Diffraction}
(Spring-Verlag, Berlin, 2002).


\bibitem{cornwall1} J. M. Cornwall, Phys. Rev. D {\bf 26}, 1453 (1982).

\bibitem{cornwall2}
J. M. Cornwall, J. Papavassiliou, Phys. Rev. D 
{\bf 40}, 3474 (1989);.

\bibitem{cornwall3}
J. Papavassiliou and J. M. Cornwall, Phys. Rev. D \textbf{44}, 1285 (1991) 

\bibitem{alkofer}
D. Binosi and J. Papavassiliou, Phys. Rept. \textbf{479}, 1 (2009).

\bibitem{amn2}
A.C. Aguilar, A. Mihara and A.A. Natale, Int. J. Mod. Phys. A \textbf{19}, 249 (2004).

\bibitem{an}
A.C. Aguilar and A.A. Natale, J. High Energy Phys. \textbf{8}, 57 (2004).

\bibitem{gribov} L.V. Gribov, E.M. Levin, and M.G. Ryskin, Phys. Rep. \textbf{100}, 1 (1983).

\bibitem{ryskin01} E.M. Levin and M.G. Ryskin, Phys. Rep. \textbf{189}, 267 (1990).









\bibitem{qcdi1} 
B. Margolis, P. Valin, M.M. Block, F. Halzen and R.S. Fletcher, 
Phys. Lett. B \textbf{213}, 221 (1988).

\bibitem{qcdi2} 
M. Block, R. Fletcher, F. Halzen, B. Margolis and P. Valin, Nucl. Phys. B (Proc. Suppl.) \textbf{12}, 238 (1990).

\bibitem{qcdi3} 
M. M. Block, F. Halzen, B. Margolis, Phys. Rev. D \textbf{45}, 839 (1992).


\bibitem{block1} 
M. M. Block, E. M. Gregores, F. Halzen, G. Pancheri, Phys. Rev. D \textbf{60}, 054024 (1999).

\bibitem{block2} 
M. M. Block, Phys. Rep. 436 (2006) 71-215.


\bibitem{wuyang}
T.T. Wu and C.N. Yang, Phys. Rev. B \textbf{137}, 708 (1965). 

\bibitem{chouyang1}
T.T. Chou and C.N. Yang, Phys. Rev. \textbf{170}, 1591 (1968). 

\bibitem{chouyang2}
T.T. Chou and C.N. Yang, Phys. Rev.  Lett. \textbf{20}, 1213 (1968).
 
\bibitem{durandlipes}
L. Durand and R. Lipes, Phys. Rev. Lett. \textbf{20}, 637 (1968).


\bibitem{georgi01} H.M. Georgi, S.L. Glashow, M.E. Machacek, and D.V. Nanopoulos, Annals of Phys. \textbf{114},
273 (1978).


\bibitem{owens01} J.F. Owens, E. Reya, and M. Gl\"uck, Phys. Rev. D \textbf{18}, 1501 (1978).


\bibitem{cutler01} R. Cutler and D. Sivers, Phys. Rev. D \textbf{17}, 196 (1978).














\bibitem{am}
R.F. \'Avila, M.J. Menon, Nucl Phys. A, \textbf{744}, 249 (2004).


\bibitem{totemslope}
K. Eggert, TOTEM Status Report and First Measurement of the Total Cross-section, 
107th LHCC Meeting, http://totem.web.cern.ch/totem/.

\bibitem{pp2pp}
S. B\"ultmann \textit{et al}. (pp2pp Collaboration), Phys. Lett. B \textbf{579}, 245 (2004).



\bibitem{pdg}
K. Nakamura et al. (Particle Data Group), J. Phys. G
\textbf{37}, 075021 (2010).

\bibitem{durham}
Durham Reaction Database, http://durpdg.dur.ac.uk/HEPDATA/REAC.

\bibitem{root} http://root.cern.ch/drupal/; http://root.cern.ch/root/html/TMinuit.html.

\bibitem{achilli}
A. Achilli, Y. Srivastava, R. Godbole, A. Grau, G. Pancheri and O. Shekhovtsova,
Total and inelastic cross-sections at LHC at $\sqrt{s}$ = 7 TeV and beyond,
arXiv:1102.1949 [hep-ph] and references therein.


\bibitem{lm} E. G. S. Luna, M. J. Menon, Phys. Lett. B {\bf 565}, 123 (2003).

\bibitem{lmm}
E. G. S. Luna, M. J. Menon, J. Montanha, Nucl.
Phys. A {\bf 745}, 104 (2004).

\bibitem{d0}
D0 Collaboration, A. Brandt, PoS DIS2010 \textbf{059} (2010).


\bibitem{kkl}
J. Ka\v{s}par V. Kundr\'at, M. Lokaj\'{\i}\v{c}ek 
and J. Proch\'azka, Nucl. Phys. B \textbf{843}, 84 (2011).

\bibitem{dl}
A. Donnachie and P. V. Landshoff, Z. Phys. C 2, 55 (1979); A. Donnachie and P. V. Landshoff, Phys. Lett. B 387, 637 (1996).

\bibitem{bl}
G. P. Lepage and S. J. Brodsky, Phys. Rev. D 22, 2157 (1980).

\bibitem{bh}
M. M. Block and F. Halzen, Phys. Rev. D \textbf{72}, 036006 (2005).


\bibitem{fm}
D.A. Fagundes and M.J. Menon, Int. J. Mod. Phys. A \textbf{26}, 3219 (2011).

\bibitem{am08}
R.F. \'Avila and M.J. Menon, Eur. Phys. J. C \textbf{54}, 555 (2008).

\bibitem{cmm}
P.A.S. Carvalho, A.F. Martini and M.J. Menon, Eur. Phys. J. C \textbf{39}, 359 (2005).

\bibitem{ap1}
J. M. Cornwall and W. S. Hou, Phys. Rev. D \textbf{34}, 585 (1986).

\bibitem{ap2}
M. Lavelle,  Phys. Rev. D \textbf{44}, 26 (1991); D. Dudal, J.A. Gracey, S. P. Sorella,
N. Vandersickel, H. Verschelde, Phys. Rev. D \textbf{78}, 065047 (2008).

\bibitem{ap3}
A. C. Aguilar and J. Papavassiliou, Eur. Phys. J. A \textbf{35}, 189 (2008).

\end{thebibliography}
\end{document}